%% file: main.tex
\documentclass[sigconf]{acmart}

\usepackage{tikz}
\usepackage{amsmath}
\usepackage{graphicx}
\usepackage{booktabs} 
\usepackage{multirow}
\usepackage{makecell}
\usepackage{listings}
\usepackage{bigstrut}
\usepackage{subcaption}
\usepackage{amsfonts}
\usepackage{xcolor}
\usepackage{xspace}
\usepackage[autolanguage]{numprint}
\usepackage{xfp}

\npthousandsep{,}
\npdecimalsign{\ensuremath{.}}

\usepackage[T1]{fontenc}
\usepackage[utf8]{inputenc}
\usepackage{ae,aecompl}

\usepackage{color}
\definecolor{red}{rgb}{1,0,0}
\definecolor{green}{rgb}{0,1,0}
\definecolor{blue}{rgb}{0,0,1}
\definecolor{cyan}{rgb}{0.4,1,1}
\definecolor{orange}{rgb}{1,0.6,0}
\definecolor{dkgreen}{rgb}{0,0.6,0}
\definecolor{dkblue}{rgb}{0.1,0.37,0.51}
\definecolor{dkred}{rgb}{0.6,0,0}
\definecolor{gray}{rgb}{0.5,0.5,0.5}
\definecolor{purple}{rgb}{0.58,0,0.82}
\definecolor{yellow}{rgb}{1,1,0}
\definecolor{dkyellow}{rgb}{1,0.9,0.45}

\usepackage[lighttt]{lmodern}
\usepackage{listingsgo}
\lstdefinelanguage{json}{
    string=[s]{"}{"},
    stringstyle=\ttfamily,
    comment=[l]{:},
    commentstyle=\color{black},
}

\usepackage[font=small,labelfont=bf]{caption}

\newif\ifDRAFT
\DRAFTtrue
\DRAFTfalse

\ifDRAFT
  
  \newcommand{\fix}[1]{\textcolor{red}{#1}}
  \newcommand{\todo}[1]{{\color{red}\bf\em TODO:\/\@#1}}
  
  \newcommand{\smw}[1]{\todo{Comment by Sam: #1}}
  \newcommand{\empirical}[1]{\textcolor{blue}{#1}}
\else
  
  \newcommand{\fix}[1]{{}}
  \newcommand{\todo}[1]{{}}
  
  \newcommand{\smw}[1]{{}}
  \newcommand{\empirical}[1]{#1}
\fi

\newcommand{\EOSStartBlock}{82152667}
\newcommand{\EOSEndBlock}{118286375}
\newcommand{\XRPStartBlock}{50399027}
\newcommand{\XRPEndBlock}{55152991}
\newcommand{\TezosStartBlock}{630709}
\newcommand{\TezosEndBlock}{932530}

\newcommand{\EOScount}{631445236}
\newcommand{\Tezoscount}{7890133}
\newcommand{\XRPcount}{271546797}

\newcommand{\point}[1]{\par\smallskip\noindent\textbf{#1.}}

\newcommand{\tezaddr}[2][\footnotesize]{{#1\href{https://www.tezos.id/accounts/#2}{\texttt{#2}}}}

\newcommand{\xrpaddr}[2][\footnotesize]{{#1\href{https://xrpscan.com/account/#2}{\texttt{#2}}}}

\newcommand{\coin}[1]{\texttt{#1}}

\newcommand{\ETHRate}{200}
\newcommand{\ToUSD}[1]{\numprint{\the\numexpr #1 * \ETHRate\relax}}

\newcommand{\startdate}{October~1,~2019\xspace}

\newcommand{\finishdate}{April~30,~2020\xspace}

\newcommand{\blockscount}[2]{\numprint{\fpeval{#2 - #1 + 1}}}
\newcommand{\tps}[2]{%
\nprounddigits{#2}%
\numprint{\fpeval{#1 / 213 / 24 / 3600}}\xspace%
}

\copyrightyear{2020} 
\acmYear{2020} 
\setcopyright{acmcopyright}\acmConference[IMC '20]{ACM Internet Measurement Conference}{October 27--29, 2020}{Virtual Event, USA}
\acmBooktitle{ACM Internet Measurement Conference (IMC '20), October 27--29, 2020, Virtual Event, USA}
\acmPrice{15.00}
\acmDOI{10.1145/3419394.3423628}
\acmISBN{978-1-4503-8138-3/20/10}


\renewcommand\footnotetextcopyrightpermission[1]{} 

\AtEndDocument{\par\leavevmode}
\begin{document}
\sloppy

\title{
Revisiting Transactional Statistics of\\ High-scalability Blockchains}

\author{Daniel Perez}
\affiliation{%
  \institution{Imperial College London}
}

\author{Jiahua Xu}
\affiliation{%
UCL CBT\\
EPFL
}

\author{Benjamin Livshits}
\affiliation{%
Imperial College London\\%
UCL CBT\\%
Brave Software}

\input{sections/0_abstract}


\maketitle

\input{sections/1_introduction}
\input{sections/2_background}
\input{sections/3_methodology}
\input{sections/4_data}
\input{sections/5_case_studies}
\input{sections/6_discussion}
\input{sections/7_related}
\input{sections/8_conclusion}
\input{sections/99_acks}

\bibliographystyle{ACM-Reference-Format}
\bibliography{references,extra}

\appendix
\section*{Appendix: Measurement Framework}

\input{sections/a_framework}

\end{document}

%% file: sections/0_abstract.tex
\begin{abstract}
Scalability has been a bottleneck for major blockchains such as Bitcoin and Ethereum.
Despite the significantly improved scalability claimed by several high-profile blockchain projects, there has been little effort to understand how their transactional throughput is being used. 
In this paper, we examine recent network traffic of three major high-scalability blockchains---EOSIO, Tezos and XRP Ledger (XRPL)---over a period of seven months.
Our analysis reveals that only a small fraction of the transactions are used for value transfer purposes. In particular, \empirical{96\%} of the transactions on EOSIO were triggered by the airdrop of a currently valueless token; on Tezos,~\empirical{76\%} of throughput was used for maintaining consensus; and over \empirical{94\%} of transactions on XRPL carried no economic value. We also identify a persisting airdrop on EOSIO as a DoS attack and detect a two-month-long spam attack on XRPL. The paper explores the different designs of the three blockchains and sheds light on how they could shape user behavior.
\end{abstract}

%% file: sections/1_introduction.tex

\section{Introduction}
\label{sec:introduction}

As the most widely-used cryptocurrency and the first application of a blockchain system, Bitcoin has been frequently criticized for its slow transactional throughput, making it hard to adopt as a payment method.
Indeed, Bitcoin is only able to process around~10 transactions per second, significantly slower than the throughput offered by centralized payment providers such as Visa, which can process over 65,000 transactions per second~\cite{visa-tps}.
Many blockchains have since been designed and developed in order to improve scalability, the most valued of these in terms of market capitalization~\cite{CoinMarketCap2020} being EOSIO~\cite{EOS}, Tezos~\cite{Goodman2014}, and XRP Ledger (XRPL)~\cite{xrp_ledger_overview}. 

Although many of these systems have existed for several years already, to the best of our knowledge, no in-depth evaluation of the actual usage of their transactional throughput has yet been performed, and it is unclear up to what point these blockchains have managed to generate economic activity.
The knowledge of both the quantity and the quality of the realized throughput is crucial for the improvement of blockchain design, and ultimately a better utilization of blockchains.
In this paper, we analyze transactions of the three blockchains listed above and seek to find out:

\begin{enumerate}
    \item[RQ1] To what extent has the alleged throughput capacity been achieved in those three blockchains?
    \item[RQ2] Can we classify transactions by analyzing their metadata and patterns?
    \item[RQ3] Who are the most active transaction initiators and what is the nature of the transaction they conducted?
    \item[RQ4] Can we reliably identify DoS and transactional spam attacks by analyzing transaction patterns?
\end{enumerate}

\point{Contributions} 
We contribute to the body of literature on blockchain in the following ways:  
\begin{enumerate}
\item We perform the first large-scale detailed analysis of transaction histories of three of the most widely-used high-throughput blockchains: EOSIO, Tezos, and XRPL.

\item We classify on-chain transactions and measure each category's respective share of the total throughput, in terms of the number of transactions and their economic volume.

\item We establish a measurement framework for assessing the quality of transactional throughput in blockchain systems.

\item We expose spamming behaviors that have inflated throughput statistics and caused network congestion.

\item We highlight the large gap between the alleged throughput capacity and the well-intended transactions being performed on those three blockchains.
\end{enumerate}
Our analysis serves as the first step towards a better understanding in the nature of user activities on high-scalability blockchains. On-chain monitoring tools can be built based on our framework to detect undesired or even malicious behavior.

\point{Summary of our findings}
Despite the advertised high throughput and the seemingly commensurate transaction volume, a large portion of on-chain traffic, including payment-related transactions, does not result in actual value transfer. 
The nature and purpose of non-payment-related activities varies significantly across blockchains.

Specifically, we observe that the current throughput is only \tps{\EOScount}{0}TPS (transactions per second) for EOSIO, \tps{\Tezoscount}{2} TPS for Tezos and \tps{\XRPcount}{0} TPS for XRPL.
We show that \empirical{96\%} of the throughput on EOSIO was used for the airdrop of a valueless token, \empirical{76\%} of transactions on the Tezos blockchain were used to maintain consensus,
and that over \empirical{94\%} of transactions on XRPL carried zero monetary value.


%% file: sections/2_background.tex
\section{Background}
\label{sec:background}

In this section, we briefly explain the fundamentals of permissionless blockchains and describe the structure of the three blockchain systems that we evaluate, highlighting their various design aspects. 

\subsection{Blockchain Basics}
In its essence, a blockchain is an append-only, decentralized database that is replicated across a number of computer nodes. 
Most blockchain systems record activities in the form of ``transactions''. 
A transaction typically contains information about its sender, its receiver, as well as the action taken, such as the transfer of an asset. 
Newly created transactions are broadcast across the network where they get validated by the participants. 
Valid transactions are grouped into data structures called \textit{blocks}, which are appended to the blockchain by referencing the most recent block.
Blocks are immutable, and state changes in the blockchain require new blocks to be produced.

Network latency and asynchrony inherent in the distributed nature of blockchains lead to a number of challenges. 
In particular, a blockchain must be able to reach consensus about the current state when the majority of participating nodes behave honestly. 
In order to resolve disagreement, a consensus protocol prescribing a set of rules is applied as part of the validation process. 


The Proof-of-Work (PoW) consensus,
introduced by Bitcoin and currently also implemented by Ethereum, requires the participant to solve a computationally expensive puzzle to create a new block. Although PoW can maintain consistency well, it is by nature very time- and energy-consuming, which limits its throughput.
To preserve security while maintaining a sufficient degree of decentralization, scalability is often sacrificed~\cite{Xie2019}.
Indeed, the rate of block creation for both Bitcoin and Ethereum is relatively slow---on average 10 minutes and 14 seconds per block, respectively---and the only way to increase the throughput is to increase the size of a single block, allowing for more transactions per block.

Another issue related to blockchain systems is the need for all participants to replicate the data.
Since blockchains are append-only, participants need to ensure that their storage capacity keeps pace with the ever-increasing size of blockchain data. 
It is therefore crucial for blockchains to be designed in such a way that the storage used increases only moderately with time.

\subsection{Consensus Mechanisms}

In response to the scalability issues related to PoW, many blockchains have developed other mechanisms to ensure consensus, which allow higher rates of block creation.

\point{Delegated Proof-of-Stake~(DPoS) in EOSIO}
\label{sec:DPoS}
EOSIO uses the Delegated Proof-of-Stake~(DPoS) protocol which was first introduced in Bitshares~\cite{bitshares}.

Users of EOSIO, stake \coin{EOS} tokens to their favored block producers (BPs) and can choose to remove their stake at any time. 
The~21 BPs with the highest stake are allowed to produce blocks whereas the rest are put on standby. 
Blocks are produced in rounds of~126~($6 \times 21$). 
The order of block production is scheduled prior to each round and must be agreed upon by at least~15 block producers~\cite{EOS}.

\point{Liquid Proof-of-Stake (LPoS) in Tezos}
For its consensus mechanism, Tezos employs another variant of Delegated Proof-of-Stake: the Liquid Proof-of-Stake~(LPoS)~\cite{Tezos2018}.
Tezos' LPoS differs from EOSIO's DPoS in that with the former, the number of consensus participants---or ``delegates''---changes dynamically~\cite{Tezos2018, Goodman2014}.
This is because any node with a minimum amount of staked assets, arbitrarily defined to be~\empirical{8,000} \coin{XTZ} (about 16,000 USD at the time of writing \cite{CoinMarketCap2020}), is allowed to become a delegate, who then has the chance to be selected as either a ``baker'' or an ``endorser''.
Each block is produced (``baked'') by one randomly selected baker, and verified (``endorsed'') by 32 randomly selected endorsers~\cite{Tezos2018}. 
The endorsements are included in the following block.



\point{XRP Ledger Consensus Protocol (XRP LCP) in XRPL}
XRPL is a distributed payment network created by \textit{Ripple Labs Inc.} in~2012 that uses the XRP ledger consensus protocol~\cite{Chase2018}.
Each user sets up its own unique node list of validators (UNL) that it will listen to during the consensus process. 
The validators determine which transactions are to be added to the ledger. 
Consensus is reached if at least~\empirical{90\%} of the validators in each ones' UNL overlap. If this condition is not met, the consensus is not assured to converge and forks can arise~\cite{Chase2018}.











\input{sections/tables/actions-overview}
\subsection{Account and Transaction Types}
In this section we describe the types of transactions that exist on the three blockchains.

\subsubsection{EOSIO}
EOSIO differentiates between system and regular accounts.
The former are built-in accounts created when the blockchain was first instantiated, and are managed by currently active BPs, while the latter can be created by anyone.
System accounts are further divided into privileged and unprivileged accounts. 
Privileged accounts, including \texttt{eosio}, \texttt{eosio.msig}, and \texttt{eosio.wrap}, can bypass authorization checks when executing a transaction \cite{EOSIO2019, Kauffman2019} (see Section~\ref{sec:DPoS}).

EOSIO system contracts, defined in \texttt{eosio.contracts}~\cite{EOSIO2020}, are held by system accounts. 
One of the most commonly used system contracts is \texttt{eosio.token}, which is designed for creating and transferring user-defined tokens~\cite{EOSIO2019}.
Regular accounts can freely design and deploy smart contracts. 

Each smart contract on EOSIO has a set of actions. 
Actions included in non-system contracts are entirely user-defined, and users have a high degree of flexibility in terms of structuring and naming the actions. 
This makes the analysis of actions challenging, as it requires understanding their true functionality on a case-by-case basis. 
While many actions have a candid name that gives away their functionality (e.g. \texttt{payout} from contract \texttt{betdicegroup}), some are less expressive (e.g. \texttt{m} from user \texttt{pptqipaelyog}).

In \autoref{tab:transaction-types-distribution}, we show different types of existing actions. 
Since actions from non-system contracts have arbitrary designs, we only examine actions that belong to system accounts for the moment, as these are already known and easier to classify. We make one exception to this and include the actions of \emph{token} contracts, as they have a standardized interface~\cite{eosio-tokens}.
Overall, we can see that token transfers account alone for more than~\empirical{96\%} of the transactions. The rest of the transactions are mostly user-defined and appear under ``Others'' in the table, while actions defined in system contracts only account for a very small percentage of the entire traffic volume.



\subsubsection{Tezos}
Tezos has two types of accounts: implicit and originated.
Implicit accounts are similar to the type of accounts found in Ethereum, generated from a public-private key pair~\cite{Wood2014}.
These accounts can produce---or ``bake''---blocks and receive stakes, but cannot be used as smart contracts.
Bakers' accounts must be implicit, to be able to produce blocks.
Originated accounts are created and managed by implicit accounts, but do not have their own private key~\cite{NomadicLabs2018a}. 
They can function as smart contracts, and can delegate voting rights to bakers' implicit accounts~\cite{Labs2018}.

``Transactions'' on Tezos are termed  ``operations.''
Operations can be roughly classified into three types: consensus related, governance related and manager operations~\cite{AmitPanghal2019}.
Consensus-related operations, as the name indicates, ensure that all participating nodes agree on one specific version of data to be recorded on the blockchain. 
Governance-related operations are used to propose and select a new set of rules for the blockchain. 
However, these events are very rare and only involve bakers, which is why these operations only represent a low percentage of the total number of transactions. 
Operations mainly consist of delegations and peer-to-peer payment transactions. 
As shown in \autoref{tab:transaction-types-distribution}, \texttt{endorsement} operations account for a vast majority,~\empirical{76\%}, of total operations.
Endorsements are performed by bakers, and a block needs a minimum of 32 endorsements for it to be accepted~\cite{NomadicLabs2018b}.


\subsubsection{XRPL}
XRPL also uses an account-based system to keep track of asset holdings. 
Accounts are identified by addresses derived from a public and private key pair. 
There are a handful of ``special addresses'' that are not derived from a key pair. 
Those addresses either serve special purposes (e.g. acting as the \coin{XRP} issuer) or exist purely for legacy reasons. 
Since a secret key is required to sign transactions, funds sent to any of these special addresses cannot be transferred out and are hence permanently lost~\cite{XRPLedger2019}.

XRPL has a large number of predefined transaction types. 
We show part of them in \autoref{tab:transaction-types-distribution}. 
The most common transaction types are \texttt{OfferCreate}, which is used to create a new order in a decentralized exchange (DEX) on the ledger, and \texttt{Payment}, which is used to transfer assets. 
There are also other types of transactions such as \texttt{OfferCancel} used to cancel a created order or \texttt{TrustSet} which is used to establish a ``trustline''~\cite{xrp_ledger_overview} with another account.

\subsection{Expected Use Cases}
\label{sec:usecase}
In this section, we describe the primary intended use cases of the three blockchains and provide a rationale for the way they are being used, to better understand the dynamics of actual transactions evaluated in Section~\ref{sec:data-analysis}.

\point{EOSIO} EOSIO was designed with the goal of high throughput and has a particularity compared to many other blockchains: there are no direct transaction fees. 
Resources such as CPU, RAM and bandwidth are rented beforehand, and there is no fixed or variable fee per transaction~\cite{EOS}.
This makes it a very attractive platform for building decentralized applications with a potentially high number of micro-payments. 
Many games, especially those with a gambling nature, have been developed using EOSIO as a payment platform.
EOSIO is also used for decentralized exchanges, where the absence of fees and the high throughput allow placing orders on-chain, unlike many decentralized exchanges on other platforms where only the settlement is performed on-chain~\cite{warren20170x}.

\point{Tezos}
Tezos was one of the first blockchains to adopt on-chain governance. 
This means that participants can vote to dynamically amend the rules of the consensus. 
A major advantage of this approach is that the blockchain can keep running without the need of hard forks, as often observed for other blockchains~\cite{byzantium-fork, dao-fork}. 
Another characteristic of Tezos is the use of a strongly typed programming language with well-defined semantics~\cite{NomadicLabs2018} for its smart contracts, which makes it easier to provide these for correctness. 
These properties make Tezos a very attractive blockchain for financial applications, such as the tokenization of assets~\cite{BTGPactual2019}.

\point{XRPL} 
Similar to EOSIO, XRPL supports the issuance, circulation, and exchange of customized tokens. 
However, in contrast to EOSIO, XRPL uses the IOU (``I owe you'') mechanism for payments. 
Specifically, any account on XRPL can issue an IOU with an arbitrary ticker~---~be it \coin{USD} or \coin{BTC}.
Thus, if Alice pays Bob~10 \coin{BTC} on XRPL, she is effectively sending an IOU of~10 \coin{BTC}, which literally means ``I (Alice) owe you (Bob)~10 \coin{BTC}''. 
Whether the \coin{BTC} represents the market value of Bitcoin depends on Alice's ability to redeem her ``debt''~\cite{XRPLedger2020}. 
This feature contributes to the high throughput on XRPL, as the speed to transfer a specific currency is no more constrained by its original blockchain-related limitations: For example, the transfer of \coin{BTC} on XRPL is not limited by the block production interval of the actual Bitcoin blockchain (typically~10 minutes to an hour to fully commit a block), and the transfer of \coin{USD} is not limited to the speed of the automated clearing house~(ACH) (around two days~\cite{Love2013}).



%% file: sections/tables/actions-overview.tex
\begin{figure*}[htbp]
  \centering
    \setlength{\tabcolsep}{1.4pt}
    \footnotesize
    \begin{tabular}{@{}lp{0.7in}rrp{1.2in}rrp{0.9in}rr@{}}
        \toprule
      & \multicolumn{3}{c}{\textbf{EOSIO}} & \multicolumn{3}{c}{\textbf{Tezos}} & \multicolumn{3}{c}{\textbf{ XRPL }} \\
    Category & Action name & \# & \% & Operation kind & \# & \% & Transaction type &  \#  & \% \\
    \cmidrule(r){1-1} \cmidrule(rl){2-4} \cmidrule(rl){5-7} \cmidrule(l){8-10}
    Peer-to-peer transactions & \textbf{Transfer} &    8,479,573,653  & \textbf{             96.2 } & Transaction &    1,941,230  &      21.4  & \textbf{Payment} &    100,328,458  & \textbf{     36.9} \\
      &   &   &   &   &   &   & EscrowFinish &                    677  &         0.0  \\
    \cmidrule(r){1-1} \cmidrule(rl){2-4} \cmidrule(rl){5-7} \cmidrule(l){8-10}
    Account actions & newaccount &                289,680  &                0.0  & Reveal &        113,915  &         0.0  & TrustSet &         3,339,620  &         1.2  \\
      & bidname &                244,248  &                0.0  & Origination &            3,159  &         1.3  & AccountSet &            150,401  &         0.1  \\
      & deposit &                243,881  &                0.0  & Activate &            2,659  &         0.0  & SignerListSet &              13,707  &         0.0  \\
      & linkauth & 148693 & 0.0 &   &   &   & SetRegularKey &                    734  &         0.0  \\
      & updateauth &                136,926  &                0.0  &   &   &   & DepositPreauth &                         3  &         0.0  \\
    \cmidrule(r){1-1} \cmidrule(rl){2-4} \cmidrule(rl){5-7} \cmidrule(l){8-10}
    Other actions & delegatebw &                684,449  &                0.0  & \textbf{Endorsement} &    6,957,612  & \textbf{     76.6} & \textbf{OfferCreate} &    160,451,595  & \textbf{     59.1} \\
      & undelegatebw &                461,320  &                0.0  & Delegation &          56,336  &         0.6  & OfferCancel &         7,259,908  &         2.7  \\
      & buyrambytes &                353,695  &                0.0  & Reveal nonce &            9,409  &         0.1  & EscrowCreate &                 1,393  &         0.0  \\
      & rentcpu &                187,878  &                0.0  & Ballot &                514  &         0.0  & EscrowCancel &                      84  &         0.0  \\
      & voteproducer &                137,713  &                0.0  & Proposals &                  90  &         0.0  & PaymentChannelClaim &                    172  &         0.0  \\
      & buyram &                  89,971  &                0.0  & Double baking evidence &                    4  &         0.0  & PaymentChannelCreate &                      33  &         0.0  \\
      & Others &        332,799,590  &                3.8  &   &   &   & EnableAmendment &                      12  &         0.0  \\
    \midrule
    \midrule
    Total &   &    8,815,351,697  &           100.0  &   &    9,084,928  &    100.0  &   &    271,546,797  &    100.0  \\
    \bottomrule
    \end{tabular}%
    \caption{Distribution of action types per blockchain.}
    \label{tab:transaction-types-distribution}%
\end{figure*}

%% file: sections/3_methodology.tex
\input{sections/tables/dataset}
\section{Methodology}
\label{sec:methodology}
In this section, we describe the methodology used to measure the transactional throughput of the selected blockchains.

\subsection{Definitions}
We first introduce important definitions used in the rest the paper.

\point{Throughput-related definitions}
When quantified, a throughput value is expressed in TPS (transactions per second).
\begin{description}
\item[Alleged Capacity] The theoretical capacity that a blockchain claims to be able to achieve
\item[Average Throughput] Average throughput recorded on the network throughout the observation period
\item[Maximum Throughput] Maximum throughput recorded on the network during the observation period
\end{description}

\point{Blockchain-related definitions}
We unify the terms that we use across the systems analyzed in this work.
We sometimes diverge from the definition provided by a particular blockchain for terminological consistency.

\begin{description}
\item[Block] Entity recorded on a blockchain and included as part of the blockchain to \emph{advance} its state. Blocks are named as such on EOSIO and Tezos, the equivalent of which on XRPL is termed  a ``ledger''.
\item[Transaction] Entity included at the top-level of a block and typically representing an atomic state transition. These are named as such on EOSIO and XRPL but are called ``operations'' in Tezos.
\item[Action] Entity included as part of the transaction and describing what the transaction should do. EOSIO and Tezos can have multiple actions per transaction. A single transaction containing multiple actions is only counted towards throughput once.
Actions are called as such in EOSIO and are the ``contents'' of an ``operation'' on Tezos. XRPL does not feature this concept and each XRPL transaction can be thought of as a single action.
\end{description}

\subsection{Measurement Framework}
We implement a framework to measure throughput on the different blockchains.
Our framework allows one to fetch historical data for all of the blockchains analyzed in this work, and to compute several statistics regarding blocks, transactions, actions and users.
Transactions and actions can be, where relevant, aggregated by time, type, sender or receivers.
Our framework is publicly available\footnote{\url{https://github.com/danhper/blockchain-analyzer}} and can easily be extended to add not only new statistics but also new blockchains.
We provide an extended overview of the framework in~Appendix.

\subsection{Data Collection}
We collect historical data on the three blockchains from \startdate to \finishdate.
We provide an overview of the characteristics of the data in~\autoref{tab:data-summary}.
We note that the numbers of transactions is not the same as in~\autoref{tab:transaction-types-distribution} as here we count only a transaction once, while in the previous table we counted all the actions included in a single transaction.

For all three of the blockchains, we first pinpoint the blocks which correspond to the start and end of our measurement period and use our framework to collect all the blocks included in this range.
Each time, we use publicly available nodes or data providers to retrieve the necessary data.

\point{EOSIO}
EOSIO nodes provide an RPC API~\cite{EOS.IO2020} which allows clients to retrieve the content of a single block, through the~\texttt{get\_block} endpoint~\cite{EOSDocs2020}.
EOSIO also has a list of block producers who usually provide a publicly accessible RPC endpoint. Out of~32 officially advertised endpoints, we shortlist~6 that have a generous rate limit with stable latency and throughput.

We collect data from block~\empirical{\numprint{\EOSStartBlock}} to block~\empirical{\numprint{\EOSEndBlock}}, or a total of~\empirical{\blockscount{\EOSStartBlock}{\EOSEndBlock}} blocks containing~\empirical{631,445,236} transactions, representing more than~\empirical{260}GB of data.

\point{Tezos}
Similar to EOSIO, Tezos full nodes provide an RPC API and some bakers make it publicly available. We measure the latency and throughput of several nodes and select the one for which we obtained the best results~\cite{tezos-ukraine}.
We obtain~\empirical{\blockscount{\TezosStartBlock}{\TezosEndBlock}} blocks containing~\empirical{7,890,133} transactions, for a total size of approximately \empirical{1.4} GB of data.

\point{XRPL}
XRPL has both an RPC API and a websocket API with similar features. Although there are no official public endpoints for XRPL, a high-availability websocket endpoint is provided by the XRP community~\cite{Wind2020}. We use the \texttt{ledger} method of the Websocket API to retrieve the data in the same way we did with EOSIO and Tezos.

In addition, we use the API provided by the ledger explorer XRP Scan~\cite{xrpscan} to retrieve account information including username and parent account.\footnote{A parent account sends initial funds to activate a new account.} 
Since large XRP users such as exchanges often have multiple accounts, this account information can be used to identify and cluster accounts.

In total, we analyze \blockscount{\XRPStartBlock}{\XRPEndBlock} blocks covering seven months of data, and containing a total of more than~\empirical{150} million transactions. The total size of the compressed data is about~\empirical{130} GB.









%% file: sections/tables/dataset.tex
\begin{figure*}[htbp]
  \setlength{\tabcolsep}{4pt}
  \centering
    \begin{tabular}{lrrrrrrrr}
    \toprule
          & \multicolumn{2}{c}{\bf Block index} & \multicolumn{1}{c}{\bf Count} & \multicolumn{1}{c}{\bf Count} & \multicolumn{1}{c}{\bf Storage} & \multicolumn{3}{c}{\bf Throughput (TPS)} \\
          & \multicolumn{1}{c}{from} & \multicolumn{1}{c}{to} & \multicolumn{1}{c}{\bf of blocks} & \multicolumn{1}{c}{\bf of transactions} & \multicolumn{1}{c}{\bf (.gzip, GB)} &  Alleged  & Max  & Average \\
\cmidrule(lr){2-3}
\cmidrule(lr){4-4}
\cmidrule(lr){5-5}
\cmidrule(lr){6-6}
\cmidrule(lr){7-9}

\bf EOSIO  & \numprint{\EOSStartBlock} & \numprint{\EOSEndBlock} & \blockscount{\EOSStartBlock}{\EOSEndBlock}& \numprint{\EOScount} & 264 & 4,000 \cite{Kramer} & 136 & \tps{\EOScount}{0} \\

\bf Tezos & \numprint{\TezosStartBlock} & \numprint{\TezosEndBlock} & \blockscount{\TezosStartBlock}{\TezosEndBlock} & \numprint{\Tezoscount}  & 1.4 & 40 \cite{lpos}  & 0.57 & \tps{\Tezoscount}{2} \\

\bf XRPL   &  \numprint{\XRPStartBlock} & \numprint{\XRPEndBlock} & \blockscount{\XRPStartBlock}{\XRPEndBlock} & \numprint{\XRPcount} & 130 & 65,000 \cite{Ripplea}& 
  56 & \tps{\XRPcount}{0} \\
\bottomrule
\end{tabular}
\caption{Characterizing the datasets for each blockchain. All measurements are performed from \startdate to \finishdate. Max throughput is the average TPS within a 6-hour interval that has the highest count of transactions. Storage size is computed with data saved as JSON Lines with one block per line and compressed using gzip level 6 of compression.}
\label{tab:data-summary}%
\end{figure*}

%% file: sections/4_data.tex
\section{Data Analysis}
\label{sec:data-analysis}
In this section, we present summary statistics and high-level illustrations of the transactions contained in the datasets of the three different blockchains.

\subsection{Transaction Overview}

\begin{figure}[tbp]
    \begin{subfigure}{\columnwidth}
        \centering
        \includegraphics[height=.27\textheight]{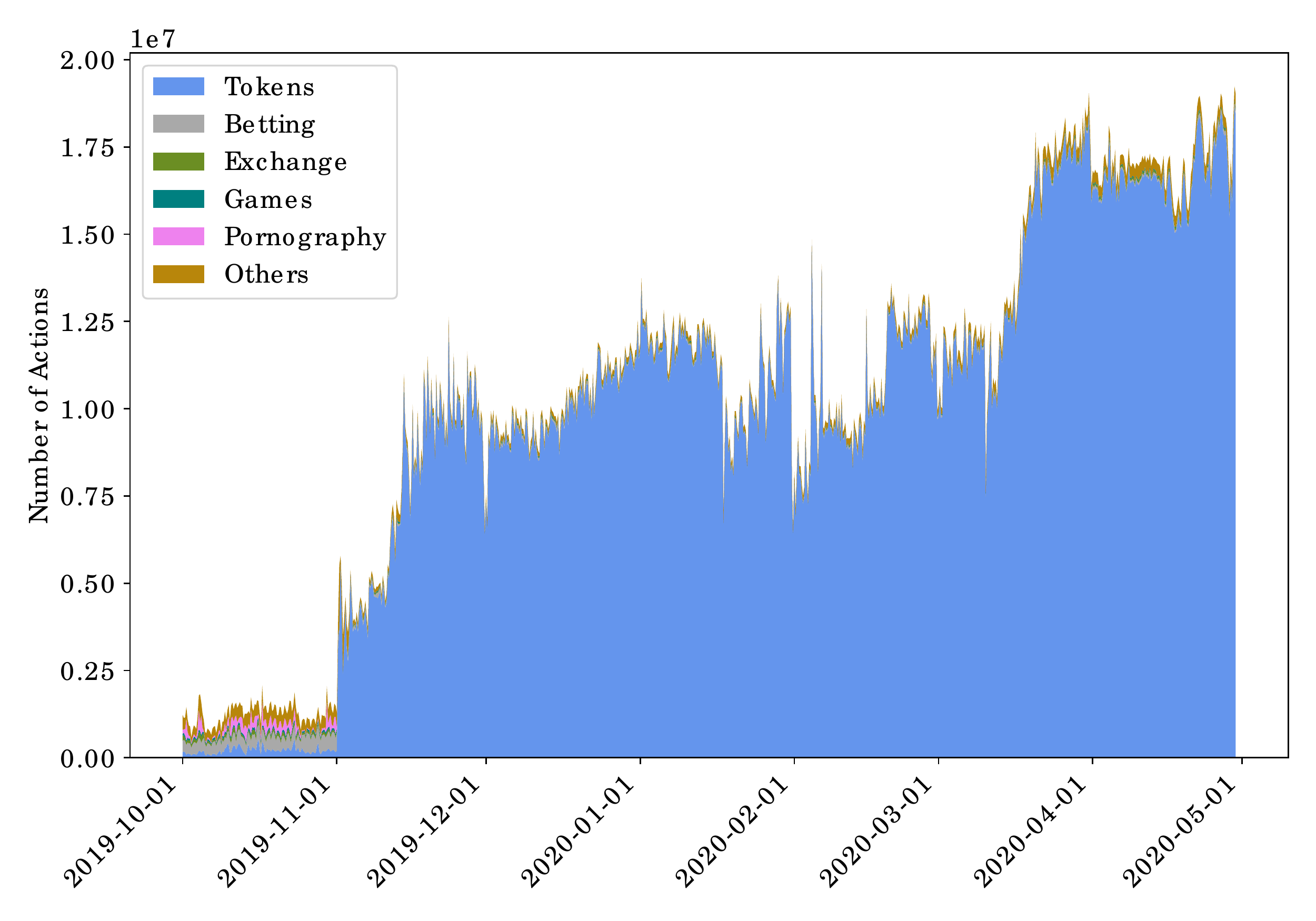}
        \caption{EOSIO throughput over time}
        \label{fig:eos-throughput-time}
    \end{subfigure}
    \begin{subfigure}{\columnwidth}
        \centering
        \includegraphics[height=.27\textheight]{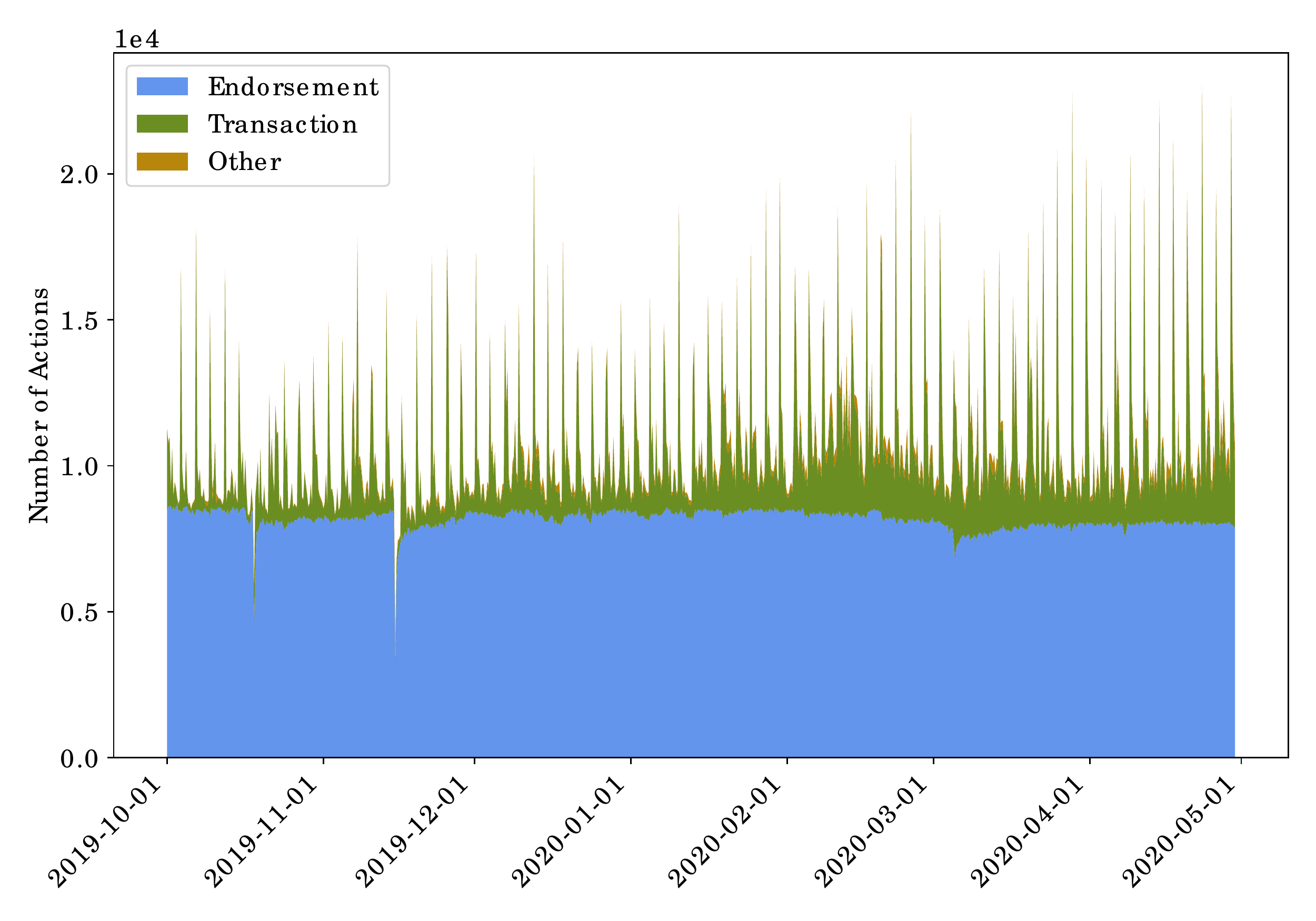}
        \caption{Tezos throughput over time}
        \label{fig:tezos-throughput-time}
    \end{subfigure}
    \begin{subfigure}{\columnwidth}
        \centering
        \includegraphics[height=.27\textheight]{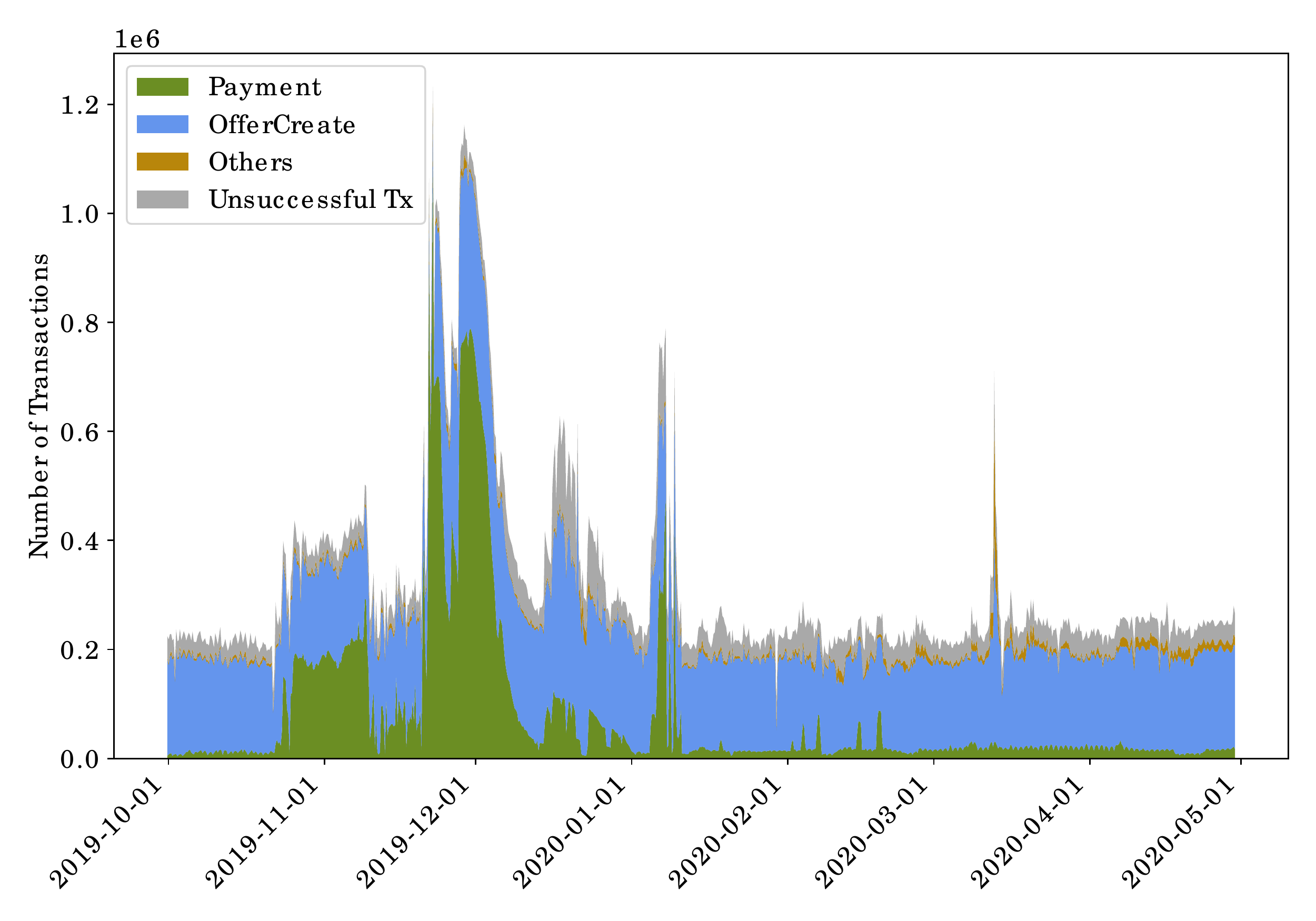}
        \caption{XRPL throughput over time}
        \label{fig:xrp-throughput-time}
    \end{subfigure}
    \caption{On-chain throughput over time, y-axis represents transaction count per 6 hours.}
    \label{fig:throughput-time}
\end{figure}

In~\autoref{fig:throughput-time}, we decompose the number of actions into different categories. XRPL and Tezos have well-defined action types, and we use the most commonly found ones to classify the throughput.
EOSIO does not have pre-defined action types: contract creators can decide on arbitrary action types.
To be able to classify the actions and understand where throughput on EOSIO is coming from, we manually label the top~\empirical{100} contracts, representing more than 99\% of the total throughput, by grouping them into different categories and assign one of the categories to each action.

\point{EOSIO}
Interestingly, there is a huge spike in the number of token actions from the November 1, 2019 onward. 
We find that this is due to a new coin called \coin{EIDOS}~\cite{Enumivo2019} giving away tokens. 
We will describe this more extensively as a case study in Section~\ref{sec:eoscase}. 
Before this peak, the number of actions on EOSIO was vastly dominated by games, in particular betting games.

\point{Tezos}
Tezos has a high number of ``endorsements''---76\%, which are used as part of the consensus protocol, and only a small fraction of the throughput are actual actions.
It is worth noting that the number of ``endorsements'' should be mostly constant regardless of the number of transactions, and that if the number of transactions were to increase enough, the trend would reverse.
We can also clearly see that Tezos has very regular spikes, with an interval of approximately two to three days each time.
These appear to be payments from bakers to stakers \cite{cryptium-labs-payout},\footnote{\url{https://twitter.com/CitezenB/status/1256147427905716224}} which can arguably be thought to be part of the consensus.
We use the TzKT API\footnote{\url{https://api.tzkt.io/}} to find account names and find that roughly \empirical{53\%} of these ``Transaction'' actions are sent by bakers and \empirical{6\%} of the are sent by the Tezos faucet~\cite{tezos-faucet}.
Endorsements and actions sent by either bakers or the faucet sums up about 87\% of the total number of actions.

\begin{figure}[tb]
    \scriptsize
    \centering
    \setlength{\tabcolsep}{2pt}
    \begin{tabular}{l l r l r}
    \toprule
    \textbf{Receiver} & \textbf{Description} & \textbf{Tx Count} & \multicolumn{2}{c}{\bf Actions}\\
         &         &    &  \bf Name   & \bf \% \\
    \midrule
      eosio.token & EOS token & 8,430,707,864 & transfer & 100.0\%\\
      \midrule
      \multirow{2}{*}{betdicetasks} & \multirow{2}{*}{Gambling game} & \multirow{2}{*}{32,804,674} & removetask & 66.65\%\\
                      & & & log & 15.71\%\\
      \midrule
      \multirow{3}{*}{whaleextrust} & \multirow{2}{*}{Decentralized} & \multirow{3}{*}{26,102,077} & verifytrade2 & 18.63\%\\
                      & & & verifytrade3 & 17.52\%\\
                      & exchange & & clearing & 16.77\%\\
      \midrule
      pptqipaelyog & Unknown & 24,109,437 & m & 93.00\%\\
      \midrule
      pornhashbaby & Pornography website & 23,677,938 & record & 99.86\%\\
    \bottomrule
    \end{tabular}
    \caption{EOSIO top applications as measured using the number of received transactions.}
    \label{tab:eos-top-applications}
    \vspace{-0.3cm}
\end{figure}

\point{XRPL}
On XRPL, both successful and unsuccessful transactions are recorded. A successfully executed transaction executes the command---such as \texttt{Payment}, \texttt{OfferCreate}, \texttt{OfferCancel}---specified by its initiator, while the only consequence of an unsuccessful transaction is the deduction of transaction fees from the transaction initiator.
Across the sample period, roughly~\empirical{one tenth} of transactions are unsuccessful (\autoref{fig:xrp-throughput-time}), with the most frequently registered errors being \texttt{PATH\_DRY} for \texttt{Payment} (insufficient liquidity, hence ``dry'', for specified payment path) and \texttt{tecUNFUNDED\_OFFER} for \texttt{OfferCreate} (no funds for the currency promised to offer by the offer creator). 

Successful transactions primarily consist of \texttt{Payment} (39.6\%) and \texttt{OfferCreate} (59.1\%) (\autoref{tab:transaction-types-distribution}). The number of \texttt{OfferCreate} is generally constant across time, but the number of \texttt{Payment} has a very high variance, with some periods containing virtually no payments and others having significant spikes. 
In Section \ref{sec:xrpcase}, we reveal why most transactions during these high-volume periods are economically meaningless. 

Except for the two spam periods, we observe that \texttt{OfferCreate} is the most common transaction type. 
Nonetheless, \texttt{OfferCreate} transactions contribute little to the total volume on XRPL.\footnote{
On May 10, 2020, for example, Ripple reported that the 24-hour XRP ledger trade volume---enabled via \texttt{OfferCreate} transactions---only accounts for \empirical{1\%} of the total ledger volume, while the payment volume---enabled via \texttt{Payment} transactions---accounts for \empirical{99\%}.
} This is because an offer expires if not fulfilled (fully or partially) before the expiry time defined by the offer creator; it can also be cancelled by its creator or superseded by a new offer. In fact, \empirical{0.2\%} of \texttt{OfferCreate} transactions resulted into an actual token exchange deal during our observation period.



 


\subsection{Transaction Patterns}
To understand better what the major sources of traffic constitute, we analyze the top accounts on EOSIO, Tezos, and XRPL, and find various transaction patterns.

\point{EOSIO}
In \autoref{tab:eos-top-applications}, we show EOSIO accounts with the highest number of received actions.
We can see that the \texttt{eosio.token} account, which is the account used to handle \coin{EOS} token transfers, is by far the most used account, and that almost all calls to this account use the \texttt{transfer} action.
Although \coin{EOS} transfers are indeed a central part of the EOSIO ecosystem, more than~\empirical{99.9}\% of the transfers shown are exclusively to and from this EIDOS account.
The second account is a betting website where all the bets are performed transparently using EOSIO.
However, around~\empirical{80\%} of the actions---\texttt{removetask} and \texttt{log}---are bookkeeping, and the actual betting-related actions such as \texttt{betrecord} represent a very low percentage of the total number of actions.
The third account is a decentralized exchange and is used to exchange different assets available on EOSIO. This exchange will be discussed in Section~\ref{sec:case-studies}.
We could not find information about the fourth account, but it is very actively sending \coin{EOS} tokens to the EIDOS account.
Finally, the last account was a pornography website which used EOSIO as a payment system. This account is still the fifth account with the highest number of received actions although the service was discontinued in November 2019 for financial reasons~\cite{hashbaby-closing}.

\input{sections/tables/tezos-top-senders}

\point{Tezos}
As Tezos neither has account names nor actions in the transactions metadata, analyzing the top receivers' accounts is less interesting, as it is very difficult to perform any type of attribution. However, we find interesting patterns from observing the top sending accounts.
Most of the top senders in Tezos seem to follow a similar pattern: Sending a small number of transactions (between~\empirical{5} and~\empirical{50}) to many different accounts. 
Another important thing to note is that all of these accounts are not contracts but regular accounts, which means that the transactions are automated by an off-chain program.
After further investigation, we find that the top address is the Tezos Faucet~\cite{tezos-faucet}. The other addresses appear to be bakers' payout addresses and the transactions are payouts to stakers~\cite{backerei}, corresponding to the peaks seen in~\autoref{fig:tezos-throughput-time}. For completeness, we include the top senders and some statistics about them in \autoref{tab:tezos-account-edges}.

\point{XRPL}
From \startdate to \finishdate, a total of~\empirical{195 thousand} accounts collectively conducted~\numprint{\fpeval{\XRPcount/1000000}} million transactions, i.e. an average of~\empirical{1.4 thousand} transactions per account during the seven-month observation period.

The distribution of the number of transactions per account is highly skewed. 
\empirical{Over one third~(71 thousand)} of the accounts have transacted only once during the entire observation period, whereas the~\empirical{35} most active accounts are responsible for half of the total traffic. 
\autoref{tab:xrpspammers} lists of the top~10 accounts by the number of conducted transactions. 
With the exception of \xrpaddr{rKLpjpCoXgLQQYQyj13zgay73rsgmzNH13} and \xrpaddr{r96HghtYDxvpHNaru1xbCQPcsHZwqiaENE}, all these accounts share suspiciously similar patterns:
\begin{enumerate}
    \item more than~98\% of their transactions are \texttt{OfferCreate};
    \item they are either descendants of an account from \href{https://www.huobi.com/}{Huobi}, a crypto exchange founded in China, or frequently transact with descendants from Huobi;
    \item they have all transacted using \coin{CNY};
    \item their payment transactions conspicuously use the same destination tag \texttt{104398}, a field that---similar to a bank reference number---exchanges and gateways use to specify which client is the beneficiary of the payment~\cite{XRPLedger2020b}.
\end{enumerate}
The aforementioned similarities, in particular the last one, signal that those accounts are controlled by the same entity, presumably with a strong connection to Huobi. 
The frequent placement of offers might come from the massive client base of the entity.

Notably, the sixth most active account, \xrpaddr{r96HghtYDxvpHNaru1xbCQPcsHZwqiaENE}, registered under the username \texttt{chineseyuan} only carried out \textit{one} successful \texttt{Payment} transaction during the observation period, while the rest of the over four million transactions failed with a \texttt{PATH\_DRY} error. Recall that failed transactions still occupy on-chain throughput. Therefore, it is evident that \texttt{chineseyuan} spammed the network.

\input{sections/tables/xrp-top-senders}

\subsection{Analysis Summary}
Here, we highlight some of the observations about the data described above. 
\begin{itemize}\itemsep=0pt
    \item Transactions on EOSIO can be roughly divided by the category of contracts they belong to. 
    Before the arrival of the \coin{EIDOS} token, approximately~\empirical{50\%} of these are transactions to betting games. The rest was split between token transfers and various forms of entertainment, such as games not involving betting as well as payments to pornography web sites. The launch of \coin{EIDOS} increased the total number of transactions more than tenfold, resulting in~\empirical{96\%} of the transactions being used for token transfers.
    
    \item The vast majority~(\empirical{76}\%) of transactions on Tezos are used by the \texttt{endorsement} operation to maintain consensus. This is due to the fact that blocks typically contain 32 endorsements~\cite{Tezos2018} and the number of transactions on the network is still low. The rest of the throughput is mainly used by transactions to transfer assets between accounts.
    
    \item \texttt{OfferCreate} and \texttt{Payment} are the two most popular transaction types on XRPL, accounting for \empirical{59.1\%} and~\empirical{36.9\%} of the total throughput, respectively. Between \startdate and October 8, 2019, before the systematic spamming periods, the fractions of \texttt{OfferCreate} and \texttt{Payment} are \empirical{79\%} and \empirical{18\%}, respectively. Overall, \empirical{one tenth} of the transactions fail.
\end{itemize}

%% file: sections/tables/tezos-top-senders.tex
\begin{figure}
    \setlength{\tabcolsep}{1.4pt}
    \footnotesize
    \begin{tabular}{@{}l r r r@{}}
    \toprule
               &            &           & \bf Avg. \# of\\
               & \bf Sent         & \bf Unique    & \bf transactions\\
    \bf Sender & \bf count & \bf receivers & \bf per receiver\\
    \midrule
    \tezaddr{tz1VwmmesDxud2BJEyDKUTV5T5VEP8tGBKGD} & 106,477 & 23,649 & 4.50\\
    \tezaddr{tz1cNARmnRRrvZgspPr2rSTUWq5xtGTuKuHY} & 105,202 & 2,096 & 50.19\\
    \tezaddr{tz1Mzpyj3Ebut8oJ38uvzm9eaZQtSTryC3Kx} & 93,448 & 93,444 & 1.00\\
    \tezaddr{tz1SiPXX4MYGNJNDsRc7n8hkvUqFzg8xqF9m} & 57,841 & 19,382 & 2.98\\
    \tezaddr{tz1acsihTQWHEnxxNz7EEsBDLMTztoZQE9SW} & 42,683 & 1,436 & 29.72\\
    \bottomrule
    \end{tabular}
    \caption{Tezos accounts with the highest number of sent transactions.}
    \label{tab:tezos-account-edges}
\end{figure}

%% file: sections/tables/xrp-top-senders.tex
\begin{figure*}[ht]
  \centering
  \footnotesize
  \renewcommand{\arraystretch}{0.6} 
    \begin{tabular}{llrrrr}
    \toprule
    \textbf{Account} & \textbf{Type} & \textbf{ Count } &       & \textbf{ TotalCount } & \textbf{\% of total throughput} \\
    \midrule
    \multirow{3}[0]{*}{\xrpaddr{r4AZpDKVoBxVcYUJCWMcqZzyWsHTteC4ZE}} & OfferCreate &  21,790,612  &   & \multicolumn{1}{r}{\multirow{3}[0]{*}{      22,082,431}} & \multirow{3}[0]{*}{8.13\%} \\
      & Others &       291,687  &   &   &  \\
      & Payment &              132  &   &   &  \\
      \midrule
    \multirow{3}[0]{*}{\xrpaddr{rQ3fNyLjbvcDaPNS4EAJY8aT9zR3uGk17c}} & OfferCreate &  21,716,850  &   & \multicolumn{1}{r}{\multirow{3}[0]{*}{      21,856,984}} & \multirow{3}[0]{*}{8.05\%} \\
      & Others &       140,088  &   &   &  \\
      & Payment &                46  &   &   &  \\
      \midrule
    \multirow{3}[0]{*}{\xrpaddr{rh3VLyj1GbQjX7eA15BwUagEhSrPHmLkSR}} & OfferCreate &  21,510,597  &   & \multicolumn{1}{r}{\multirow{3}[0]{*}{      21,541,929}} & \multirow{3}[0]{*}{7.93\%} \\
      & Others &         31,295  &   &   &  \\
      & Payment &                37  &   &   &  \\
      \midrule
    \multirow{3}[0]{*}{\xrpaddr{r4dgY6Mzob3NVq8CFYdEiPnXKboRScsXRu}} & OfferCreate &  21,474,131  &   & \multicolumn{1}{r}{\multirow{3}[0]{*}{      21,504,135}} & \multirow{3}[0]{*}{7.92\%} \\
      & Others &         29,841  &   &   &  \\
      & Payment &              163  &   &   &  \\
      \midrule
    \xrpaddr{rKLpjpCoXgLQQYQyj13zgay73rsgmzNH13} & Payment &    4,493,754  &   & \multicolumn{1}{r}{        4,493,754 } & 1.65\% \\
    \midrule
    \xrpaddr{r96HghtYDxvpHNaru1xbCQPcsHZwqiaENE} & Payment &    4,488,127  &   & \multicolumn{1}{r}{        4,488,127 } & 1.65\% \\
    \midrule
    \multirow{2}[0]{*}{\xrpaddr{rBW8YPFaQ8WhHUy3WyKJG3mfnTGUkuw86q}} & OfferCreate &    4,474,481  &   & \multicolumn{1}{r}{\multirow{2}[0]{*}{        4,475,448}} & \multirow{2}[0]{*}{1.65\%} \\
      & Others &              967  &   &   &  \\
      \midrule
    \multirow{2}[0]{*}{\xrpaddr{rDzTZxa7NwD9vmNf5dvTbW4FQDNSRsfPv6}} & OfferCreate &    4,472,749  &   & \multicolumn{1}{r}{\multirow{2}[0]{*}{        4,473,792}} & \multirow{2}[0]{*}{1.65\%} \\
      & Others &           1,043  &   &   &  \\
      \midrule
    \multirow{3}[0]{*}{\xrpaddr{rV2XRbZtsGwvpRptf3WaNyfgnuBpt64ca}} & OfferCreate &    4,470,525  &   & \multirow{3}[0]{*}{        4,471,578} & \multirow{3}[0]{*}{1.65\%} \\
      & Others &              977  &   &   &  \\
      & Payment &                76  &   &   &  \\
      \midrule
    \multirow{3}[0]{*}{\xrpaddr{rwchA2b36zu2r6CJfEMzPLQ1cmciKFcw9t}} & OfferCreate &    4,470,528  &   & \multirow{3}[0]{*}{        4,471,551} & \multirow{3}[0]{*}{1.65\%} \\
      & Others &           1,008  &   &   &  \\
      & Payment &                15  &   &   &  \\
    \bottomrule
    \end{tabular}%
  \caption{XRPL accounts with the highest number of transactions.}
  \label{tab:xrpspammers}%
\end{figure*}

%% file: sections/5_case_studies.tex
\section{Case Studies}
\label{sec:case-studies}

In this section, we present several case studies of how the transaction throughput on the three blockchains is used in practice,  for both legitimate and less legitimate purposes.

\subsection{Inutile Transactions on EOSIO}
\label{sec:eoscase}
\point{Exchange Wash-trading} 
We investigate WhaleEx, who claims to be the largest decentralized exchange~(DEX) on EOSIO in terms of daily active users~\cite{WhaleEx2020}. 
As shown in~\autoref{tab:eos-top-applications}, the most frequently-used action of the WhaleEx contract are \texttt{verifytrade2} and \texttt{verifytrade3}, with a combined total of~\empirical{9,437,393} calls over the seven months observational period, which corresponds to approximately \empirical{one action every two seconds}. These actions are executed when a buy offer and a sell offer match each other and signals a settled trade.



Firstly, and most obviously, we notice that in more than~\empirical{75\%} of the trades, the buyer and the seller are \emph{the same}. 
This means that no asset is transferred at the end of the action. 
Furthermore, the transaction fees for both the buyer and the seller are~0, which means that such a transaction is achieving absolutely nothing else than \emph{artificially} increasing the service statistics, i.e. wash-trading.

Further investigation reveals that accounts involved in the trades that are signaled by either \texttt{verifytrade2} or \texttt{verifytrade3} are highly concentrated: the top~5 accounts, as either a ``seller'' or a ``buyer'', are associated with over~\empirical{78\%} of the trades. 
We compute the percentage of such transactions for the top~5 accounts and find that each of these accounts acts simultaneously as both seller and buyer in more than~\empirical{88\%} of the transactions they are associated with. 
This means that the \emph{vast majority} of transactions of the top~5 accounts represent wash-trading.

Next, we analyze the amount of funds that has been moved, i.e. the difference between the total amount of cryptocurrency sent and received by the same account.
For the most active account, we find that only~\empirical{one} of the~\empirical{4} currencies has a balance change of over~\empirical{0.3\%}. 
The second most frequently used account has a similar transaction pattern, with only \empirical{2} out of the~\empirical{32} currencies traded showing a balance change larger than~\empirical{0.6\%}.
The rest of the top accounts all follow a very similar trend, with almost all the traded currencies having almost the same sent and received amounts.


\point{Boomerang transactions} 
As shown in~\autoref{fig:eos-throughput-time}, there was a very sharp increase of activity on EOSIO after November~1,~2019. After investigating, we find that this increase is due to the airdrop of a new coin called~\coin{EIDOS}~\cite{Enumivo2019}.

The token distribution works as follows: Users send any amount of \coin{EOS} to the EIDOS contract address, the EIDOS contract sends the \coin{EOS} amount back to the sender and also sends 0.01\% of the \coin{EIDOS} tokens it holds. This creates a ``boomerang'' transaction for the \coin{EOS} token and a transaction to send the \coin{EIDOS} token. The tokens can then be traded for \coin{USDT} (Tether) which can in turn be converted to other currencies. There are no transaction fees on EOSIO and users can execute transactions freely within the limits of their rented CPU capacity. Therefore, this scheme incentivizes users with idle CPU resources on EOSIO to send transactions to this address, creating a large increase in the numbers of transactions.

Soon after the launch of this coin, the price of CPU usage on EOSIO spiked by~10,000\% and the network entered a congestion mode. In a normal mode, users can consume more CPU than they staked for, but when the network is in congestion mode, they can only consume the amount staked. Although this is how the network is supposed to behave, it is problematic if it lasts for a non-negligible period of time. For example, \coin{EOS} is used for games where many users make a small number of transactions without staking CPU. When the network enters congestion mode for a long period of time, these users cannot continue to play unless they actively stake \coin{EOS} for CPU. 
This has caused some services to threaten with their migration to another blockchain~\cite{EarnBet2019EOSNotice}.

The coin seems to be operated by an entity called Enumivo but there is very scarce information about what service it provides. Given the very hostile tone in communications\footnote{\url{https://twitter.com/enumivo/status/1193353931797057536}}, it is likely that the creator indeed intended to congest the EOSIO network. Furthermore, the entity behind the \coin{EIDOS} token seems to be willing to launch a ``sidechain'' of EOSIO~\cite{yas-network}.

\begin{figure*}[tb]
    \centering
  \begin{subfigure}{0.323\textwidth}
    \centering
    \includegraphics[width=\columnwidth]{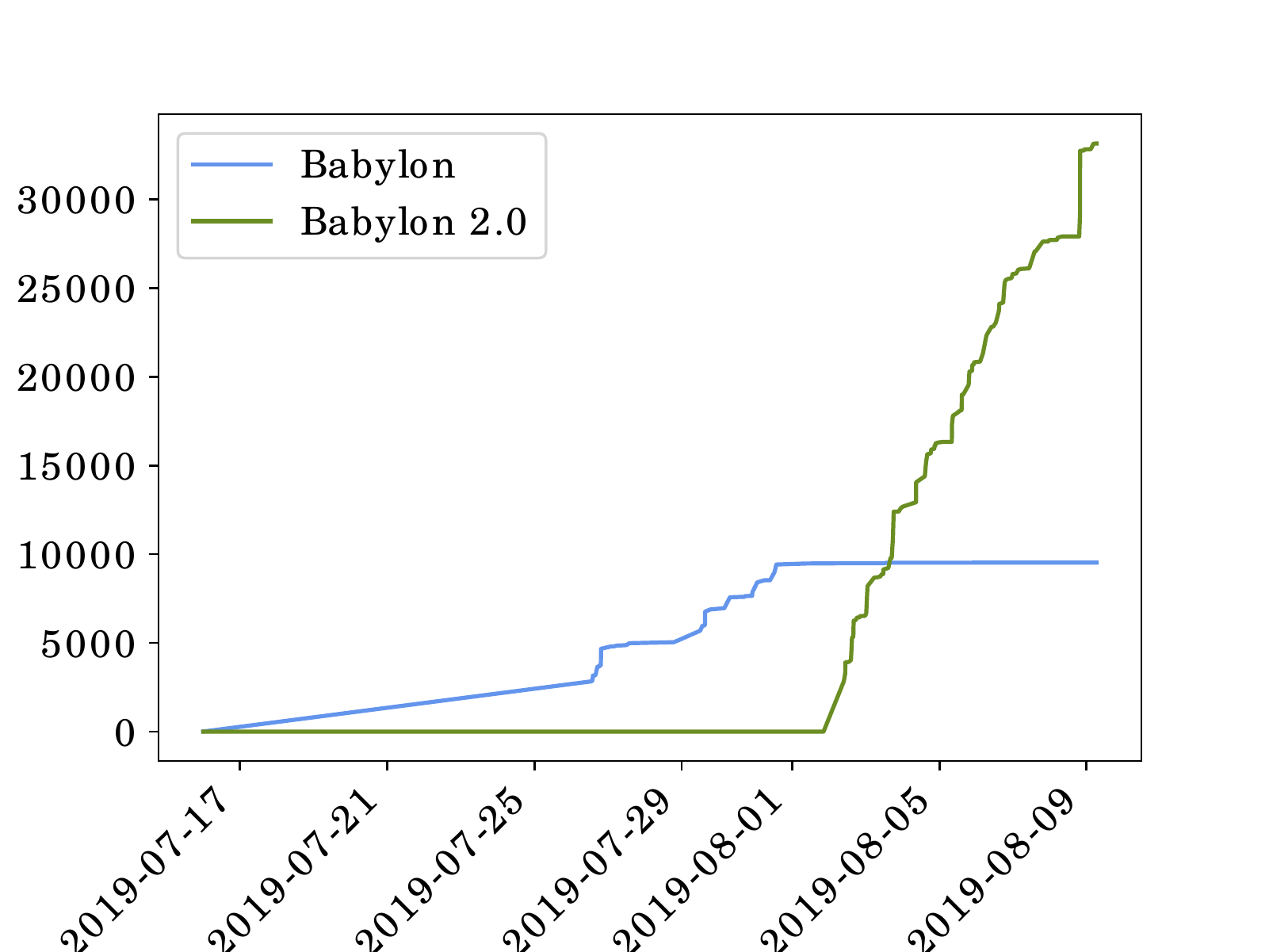}
    \caption{Proposal period}
  \end{subfigure}
  \begin{subfigure}{0.323\textwidth}
    \centering
    \includegraphics[width=\columnwidth]{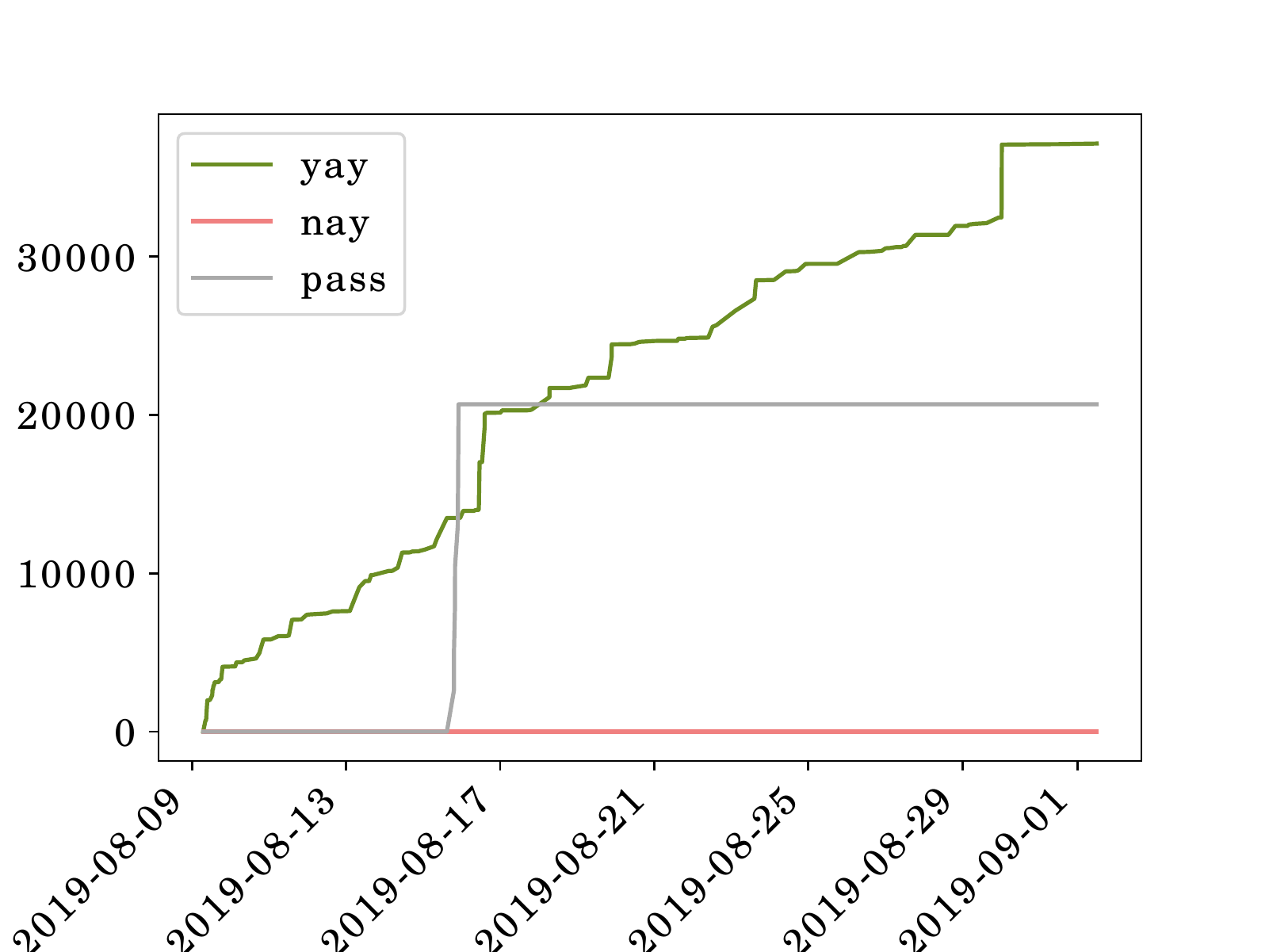}
    \caption{Exploration period}
  \end{subfigure}
  \begin{subfigure}{0.323\textwidth}
    \centering
    \includegraphics[width=\columnwidth]{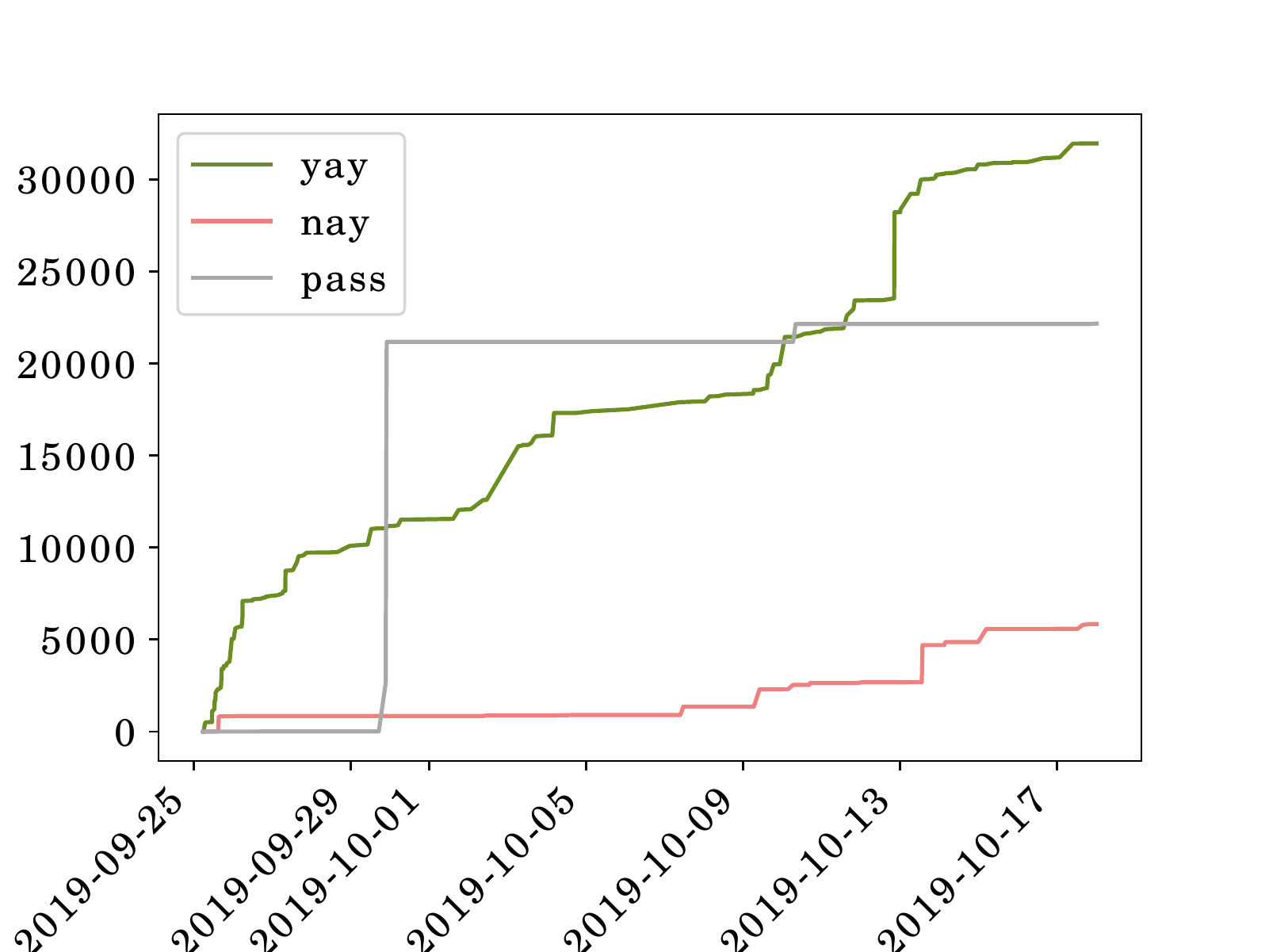}
    \caption{Promotion period}
  \end{subfigure}
  \caption{Tezos Babylon on-chain amendment voting process.}
  \label{fig:tezos-votes}
\end{figure*}

\point{Summary}
One of the major selling points of EOSIO is its absence of transaction fees for most users. Although this clearly provides advantages for users, it can also result in spamming behaviors, as observed in this section. The fee-free transaction environment encourages market manipulation such as the WhaleEx wash-trading; moreover, it has also back-fired with the \coin{EIDOS} token, as the network had to enter congestion mode and users have to stake an amount much higher than transaction fees in the Bitcoin network~\cite{EarnBet2019EOSNotice}.

\subsection{Governance Transactions on Tezos}
One of the main particularities of Tezos, compared to other blockchains, is its on-chain governance and self-amendment abilities. Given that only~\emph{bakers} are allowed to send such transactions and that they can only perform a limited number of actions within a certain time frame, governance-related transactions represent only a very small fraction of the total number of transactions: merely~\empirical{604} within our observation period. However, given that this type of transaction is rather unique and has, to the best of our knowledge, not been researched before, we analyze how the different phases of the governance process are executed in practice.

\point{Tezos voting periods} 
Tezos voting is divided into four periods, each lasting around~\empirical{23} days~\cite{Goodman2014}. During the first period, the proposal period, bakers are allowed to propose an amendment in the form of source code to be deployed as the new protocol for Tezos.
At the end of this period, the proposal with the highest number of bakers' votes is selected for the next period: The exploration period. During the exploration period, the bakers either choose to approve, refuse or abstain on voting on the proposal. If the quorum and the minimum positive votes---both thresholds are dynamically adjusted based on past participation---is reached, the proposal enters the testing period. During the testing period, the proposal is deployed on a testing network, without affecting the main network. Finally, the last period is the promotion vote period, which works in the same way as the exploration period but if successful, the new protocol is deployed as the new main network.

\point{Analyzing Tezos Voting} 
To investigate the entire voting process in Tezos, we collect extra data associated with a recent amendment called Babylon~2.0~\cite{CryptiumLabs2019}, which was proposed on August~2,~2019 and promoted to the main network on October~18,~2019. 
We show the evolution of the votes during the different voting phases in~\autoref{fig:tezos-votes}.

During the proposal period, a first proposal, ``Babylon'', was submitted and slowly accumulated votes. During this phase, the authors of Babylon received feedback from involved parties and released an updated protocol, Babylon~2.0. Votes can be placed on multiple proposals which is why the number of previous votes on Babylon did not decrease. At the end of the vote, the participation was roughly~\empirical{49\%}. It is worth noting that, although in practice any baker can propose an amendment to the network, from the creation of the Tezos blockchain up until the time of this writing, only Cryptium Labs and Nomadic Labs, who are both supported by the Tezos Foundation, have made successful proposals.

During the exploration period, participants can vote ``yay'' to support the proposal, ``nay'' to reject it, or ``pass'' to explicitly abstain from voting. No negative votes were cast during this period and the only abstention was from the Tezos Foundation, whose policy is to always abstain to leave the decision to the community. This phase had a participation of over~\empirical{81\%}, significantly higher than for the previous round. This can be explained by the fact that explicit abstention counts as participation, while there is no way to explicitly abstain in the proposal phase.

Finally, after the testing period during which the proposal was deployed and tested on a testnet, the promotion period started. The trend was mostly similar to what was observed in the exploration period, but the number of votes against the proposals increased from~\empirical{0} to~\empirical{15\%}, as some bakers encountered trouble during the testing period due to changes in the transaction format that led to breaking components~\cite{ObsidianSys}.

\point{Improvement potential on voting mechanism}
There are currently four periods in the Tezos voting system. 
First, participants can submit proposals, then they decide whether to try the elected proposal on a testing network and finally whether to amend the main network using the proposal. 
However, at the time of writing, in every exploration period seen, proposals have always received more than~99\% approval during the exploration period.
With the only exception where more than~99\% of rejections were received~\cite{tezos-vote-reject} during the exploration period, the participation during the proposal period was below 1\%.
This shows that proposals selected by a large enough number of participants are almost unanimously approved in the exploration period.
Although the current voting scheme could be useful in the future, we believe this shows that in the current state of the network, the proposal and exploration periods could be merged.
This would allow a reduction in the time until amendments ship to the main network without compromising the functionality or security of the network.

\subsection{Zero-value Transactions on XRPL}
\label{sec:xrpcase}

\point{Payments with zero-value tokens}
As described in Section~\ref{sec:usecase}, XRPL offers autonomy in currency issuance. On the flip side, this means that it is easy to generate seemingly high-value, but in effect valueless and useless transactions. 
Currencies bearing the same ticker issued by different accounts can have drastically differing valuations due to the varying level of trust in their issuers and the redeemability of their IOU tokens, which has in the past caused confusion among less informed users.\footnote{\url{https://twitter.com/Lord_of_Crypto/status/965344062084497408}}

In fact, the only currency whose value is recognized outside of XRPL is its native currency \coin{XRP}, which is also the only currency that cannot be transferred in the form of IOUs. 
Non-native currencies can be exchanged with each other or to \coin{XRP} via decentralized exchanges (DEX) on the ledger. 
Therefore, a reliable way of evaluating a currency by a certain issuer is to look up its exchange rate against \coin{XRP}. 
Normally, only IOU tokens issued by featured XRPL gateways are deemed valuable; in contrast, tokens issued by random accounts are most likely to be deemed worthless. 
For example, the value of \coin{BTC} IOUs from various issuer accounts could range from~0 to~36,050~\coin{XRP} (\autoref{fig:xchangeissuer}).

\begin{figure}[tb]
     \centering
     \begin{subfigure}[b]{\linewidth}
     \footnotesize
         \centering
         \setlength{\tabcolsep}{2.7pt}
            \begin{tabular}{llr}
            \toprule
            Issuer name        & Issuer account                                       & Rate \\
            \midrule
            Bitstamp        & \xrpaddr{rvYAfWj5gh67oV6fW32ZzP3Aw4Eubs59B}       & 36,050 \\
            Gatehub Fifth   & \xrpaddr{rchGBxcD1A1C2tdxF6papQYZ8kjRKMYcL}       & 35,817 \\
            BTC 2 Ripple    & \xrpaddr{rMwjYedjc7qqtKYVLiAccJSmCwih4LnE2q}      & 409        \\
            \emph{not registered}   & \xrpaddr{r3fFaoqaJN1wwN68fsMAt4QkRuXkEjB3W4}                 & 1        \\
            \emph{not registered}    & \xrpaddr{rpJZ5WyotdphojwMLxCr2prhULvG3Voe3X}                 & 0        \\
            \bottomrule
            \end{tabular}
         \caption{Rates (in \coin{XRP}) of \coin{BTC} IOUs issued by exemplary accounts in demonstration of the wide rate range. Each rate value is the average exchange rate of the issuer-specific \coin{BTC} IOU tokens. 
         Data retrieved through \url{https://data.ripple.com/v2/exchange_rates/BTC+{issuer_address}/XRP?date=2020-01-01T00:00:00Z&period=30day} \cite{XRPLedger2020a}.}
         \label{fig:xchangeissuer}
     \end{subfigure}
     \vskip 5pt
     \begin{subfigure}[b]{\linewidth}
     \footnotesize
         \centering
            \begin{tabular}{llr}
            \toprule
            Date        & Seller account of \coin{BTC} IOU & Rate \\
            \midrule
            2019-12-14        & \xrpaddr{rHVsygEmrjSjafqFxn6dqJWHCdAPE74Zun}       & 30,500 \\
            2020-01-09   & \xrpaddr{rU6m5F9c1eWGKBdLMy1evRwk34HuVc18Wg}       & 1 \\
            2020-01-09    & \xrpaddr{rU6m5F9c1eWGKBdLMy1evRwk34HuVc18Wg}      & 0.1        \\
            \bottomrule
            \end{tabular}
         \caption{Rate (in \coin{XRP}) of \coin{BTC} IOUs issued by \xrpaddr{rKRNtZzfrkTwE4ggqXbmfgoy57RBJYS7TS} at different time. In all the three exchange transactions, the account that buys the \coin{BTC} IOU against \coin{XRP} is \xrpaddr{rMyronEjVcAdqUvhzx4MaBDwBPSPCrDHYm}}.
         \label{fig:xchangetime}
     \end{subfigure}
     \vspace{-1cm}
        \caption{Rate (in \coin{XRP}) of \coin{BTC} IOUs on XRPL.}
        \label{tab:btcrate}
\end{figure}

The ledger experienced two waves of abnormally high traffic in the form of \texttt{Payment} transactions in late~2019, the first between the end of October and the beginning of November, the second---at a higher level---between the end of November and the beginning of December (\autoref{fig:xrp-throughput-time}). 
The culprit behind the increased traffic is \xrpaddr{rpJZ5WyotdphojwMLxCr2prhULvG3Voe3X}, an account activated on October~9,~2019 which itself managed to activate~5,020 new accounts within one week with a total of~1 million \coin{XRP} (roughly 250,000 USD), only to have them perform meaningless transactions between each other, wasting money on transaction fees. 
The behavior triggered a heated debate in the XRP community where a member claimed that the traffic imposed such a burden to their validator that it had to be disconnected~\cite{Tulo2019}.

\begin{figure}[tb]
    \begin{subfigure}{\columnwidth}
    \includegraphics[
    width=\linewidth,
    trim = {0, 20, 0, 10}, clip
    ]{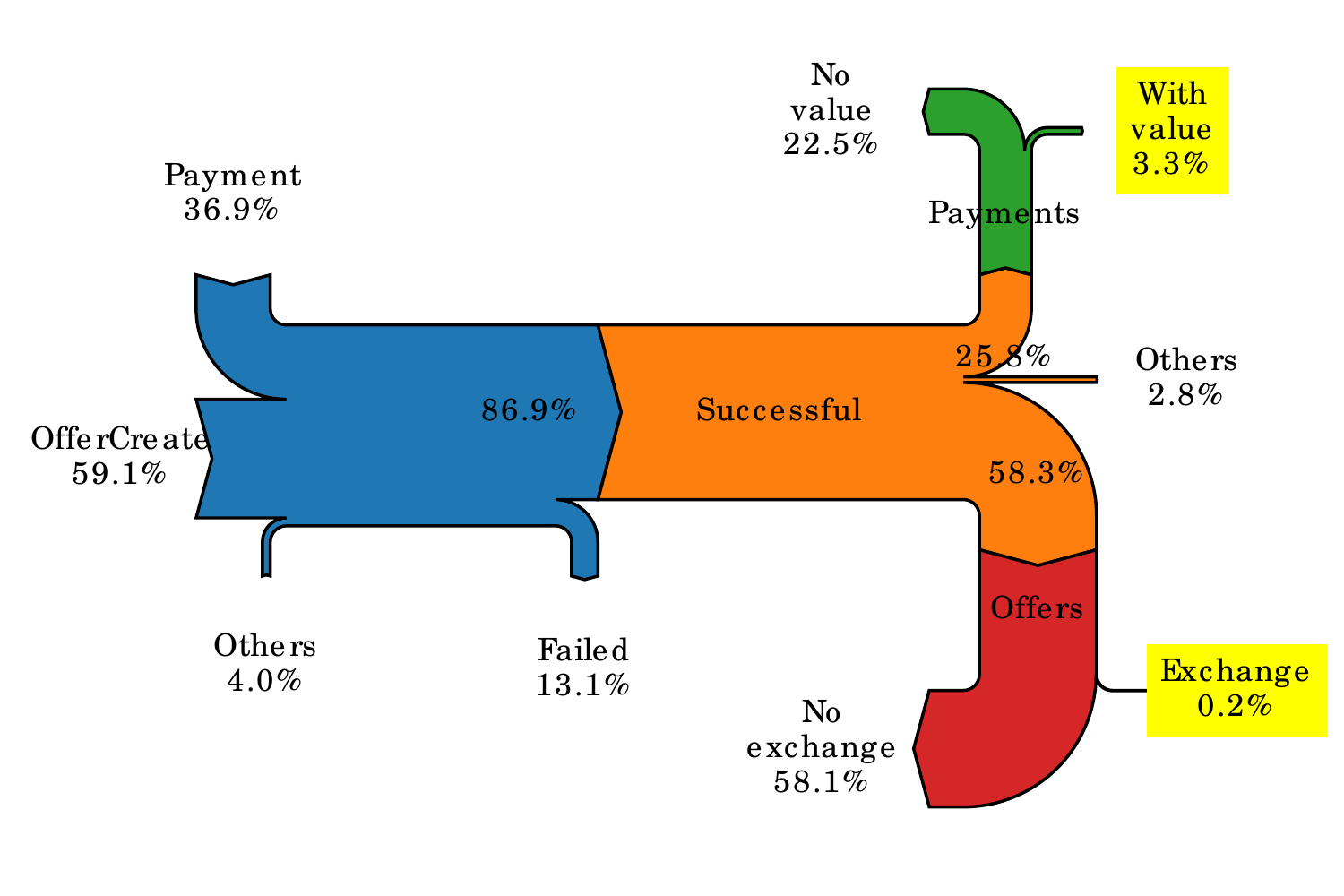}
    \caption{Observation period: \startdate to \finishdate.}
    \label{fig:xrpsakeyfull}
    \end{subfigure}
    \begin{subfigure}{\columnwidth}
    \includegraphics[
    width=\linewidth,
    trim = {0, 20, 0, 10}, clip
    ]{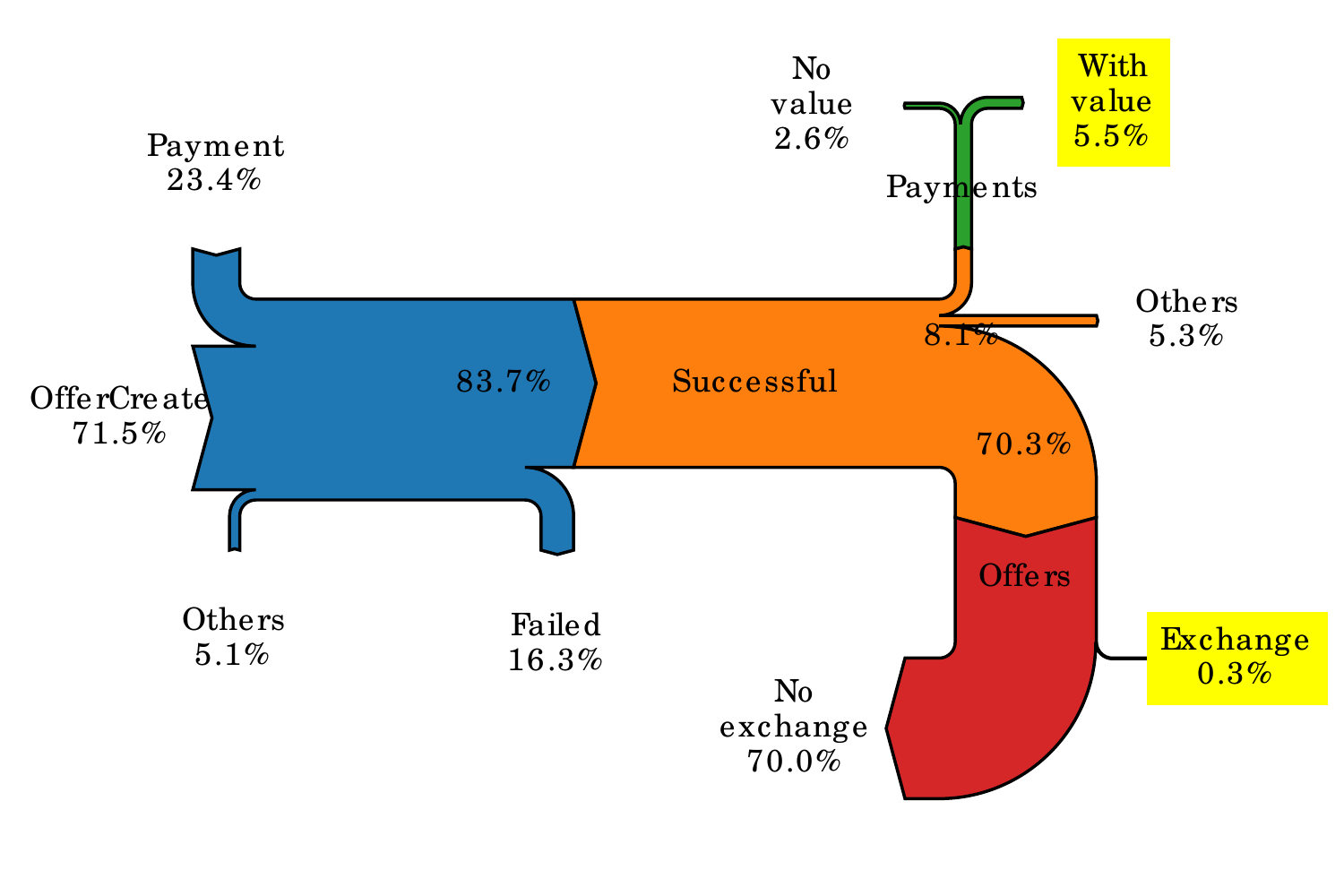}
    \caption{Observation period: February 1, 2020 to \finishdate, during which the throughput was not polluted by systematic \texttt{Payment} spams.}
    \label{fig:xrpsakeysel}
    \end{subfigure}
    \caption{XRPL throughput by transaction type, success and value transferred.
    \colorbox{yellow}{Highlighted} transactions carry economic value.}
\end{figure}

Ripple suspected it to be ``an attempt to spam the ledger'' with little impact on the network.\footnote{\url{https://twitter.com/nbougalis/status/1198670099160322048}} 
However, large exchanges such as Binance suffered from temporary \coin{XRP} withdrawal failures, who cited the XRP network congestion as the cause~\cite{AtoZMarkets2019}. 
It remains something of a mystery how such an expensive form of ``spam'' benefited its originators.

The payment transactions from the spam did not carry any value, since they involved transferring \coin{BTC} IOU tokens unacceptable outside of the spammer's network. 

To quantify true value-transferring \texttt{Payment} transactions, we retrieve the exchange rate with respect to \coin{XRP} of all the issuer-specific tokens that were transferred between \startdate and \finishdate. 
Only~\empirical{12.8\%} (3.3\%/25.8\%) of all successful \texttt{Payment} transactions involve tokens with a positive \coin{XRP} rate (\autoref{fig:xrpsakeyfull}). 

To obtain a picture of throughput usage uncontaminated by systematic spam, we re-examine the transaction data from February 1, 2020 to \finishdate. During this period, 67.9\% successful \texttt{Payment} transactions led to value transfer (\autoref{fig:xrpsakeysel}). Nevertheless, the value-carrying share of total throughput remains under \empirical{6\%}, since successful \texttt{Payment} transactions only account for a small fraction (\empirical{8.1\%}) of the overall traffic and the majority (\empirical{97.9\%}) of \texttt{OfferCreate} transactions eventually becomes void.





\begin{figure}
    \centering
    \includegraphics[width=\linewidth,
    trim = {60, 50, 62, 50}, clip]{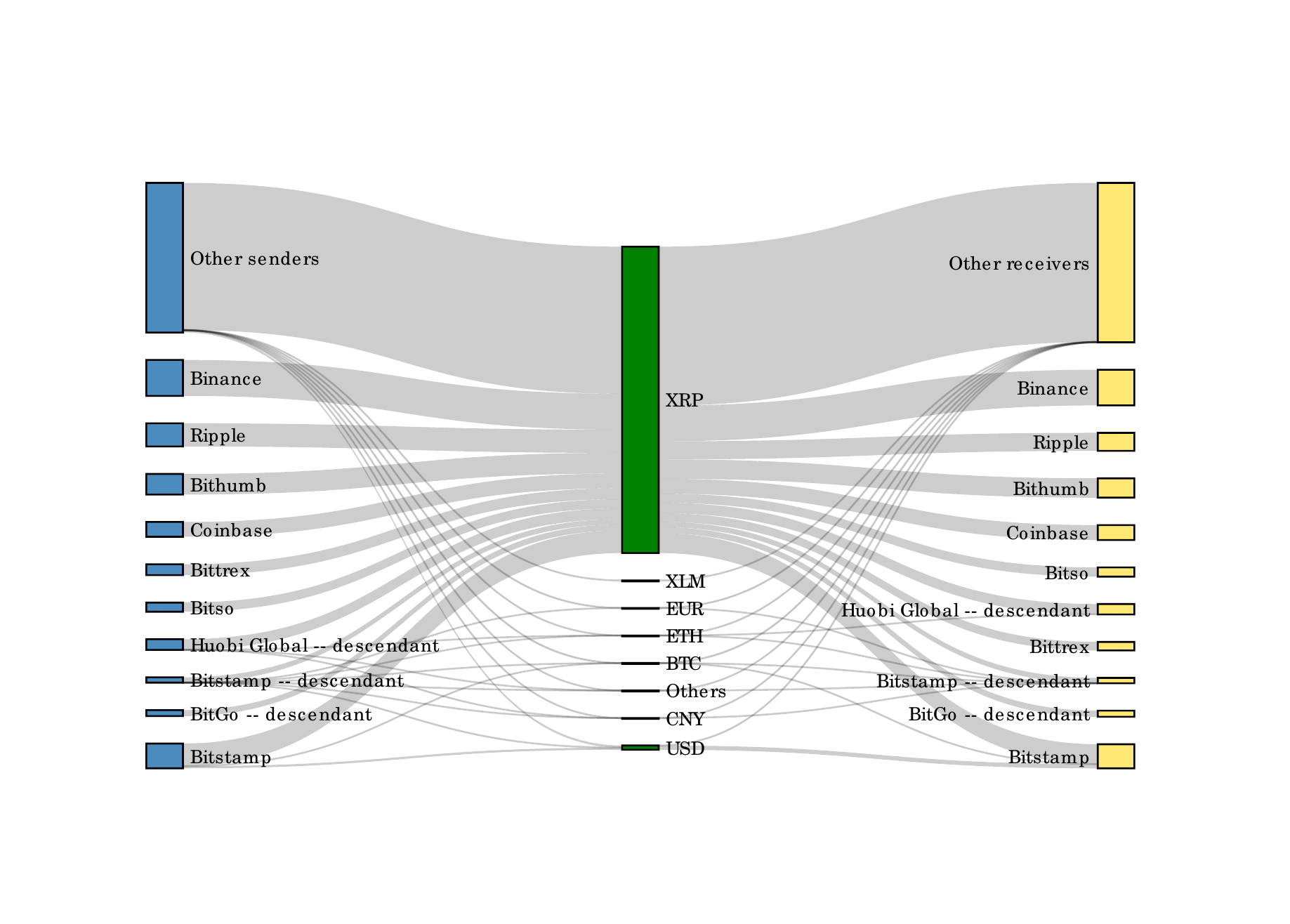}
    \tiny{
    \textcolor{dkblue}{$\blacksquare$} Sender
    \qquad
    \textcolor{dkgreen}{$\blacksquare$} Currency
    \qquad
    \textcolor{dkyellow}{$\blacksquare$} Receiver
    }
    \caption{Value flow on the XRP ledger between \startdate and \finishdate. The bandwidth of each flow represents the magnitude of aggregate value transferred denominated in \coin{XRP}. Only \texttt{Payment} transactions are included.}
    \label{fig:xrpcf}
\end{figure}

In \autoref{fig:xrpcf}, we show the top senders and receivers of value-carrying \texttt{Payment} transactions, as well as the most popular currencies being transferred. 
To cluster accounts, we rely on usernames as the identifier, as one entity can have multiple addresses under a given user name (e.g. \texttt{Binance}, \texttt{Coinbase}). 
For accounts with no registered username, we use their parent's username, if available, plus the suffix ``descendant'' as their identifier.

As one might expect, \coin{XRP} is by far the most used currency on the ledger in terms of payment volume:~\empirical{125 billion} \coin{XRP} for seven months, or~\empirical{586 million} \coin{XRP} per day. 

The top~10 senders cover~\empirical{53\%} of this volume, while the top~10 receivers are the beneficiaries of~\empirical{50\%} of the volume. 
Payments from Ripple alone account for \empirical{7\%~(9 billion \coin{XRP})} of the \coin{XRP} volume, largely due to transactions associated with the monthly release of one billion \coin{XRP} from escrows. 
While the \coin{XRP} release itself is captured through \texttt{EscrowCreate} transactions,~90\% of the released funds were unused and returned to escrows for future release~\cite{TeamRipple} through \texttt{Payment} transactions.
All other top accounts presented are held either by exchanges, or, in rare cases, by accounts that were opened by an exchange. 
\texttt{Binance} appears to be the most avid \coin{XRP} user, sending~\empirical{15.2 billion} and receiving~\empirical{14.5 billion} \coin{XRP} during the observation period.

The most popular IOU tokens for fiat currencies include \coin{USD}, \coin{EUR} and \coin{CNY} (\autoref{fig:xrpcf}). 
Specifically,~\empirical{328} million \coin{USD},~\empirical{8} million \coin{EUR} and~\empirical{19} million \coin{CNY} issued had positive exchange rates against \coin{XRP}. 
The average on-ledger exchange rates of those three fiat currency tokens, irrespective of their issuers, were~\empirical{5.4} \coin{XRP}/\coin{USD},~\empirical{5.5} \coin{XRP}/\coin{EUR} and~~\empirical{0.7} \coin{XRP}/\coin{CNY}, largely in accordance with the off-ledger exchange rates.\footnote{\url{https://finance.yahoo.com/}}

\point{Fulfilled offers with zero-value tokens}
We found a series of conspicuous payment transactions with the aggregate transfer of~\empirical{360,222} \coin{BTC} IOU, issued by \xrpaddr{rKRNtZzfrkTwE4ggqXbmfgoy57RBJYS7TS}, an account activated by Liquid (\href{https://www.liquid.com/}{liquid.com}), from the issuer itself to \xrpaddr{rMyronEjVcAdqUvhzx4MaBDwBPSPCrDHYm}, an account activated by uphold (\href{https://uphold.com/}{uphold.com}). 
The \coin{BTC} IOU token was exchanged at~\empirical{30,500} \coin{XRP}, resulting in a valuation of~\empirical{11} billion \coin{XRP} of those payments. We examine the legitimacy of the exchange rates in the next step.

The issuer is not the only factor behind the value of an IOU token. 
Even IOU tokens for the same currency from the same issuer can at times exhibit vastly different rates. 
\autoref{fig:xchangetime} shows an example where the \coin{BTC} IOU from the same issuer \xrpaddr{rKRNtZzfrkTwE4ggqXbmfgoy57RBJYS7TS} was traded at~\empirical{30,500} \coin{XRP} in December~2019 but then declined to~0.1~\coin{XRP} within a month.

The three exchange instances in \autoref{fig:xchangeissuer} were \texttt{OfferCreate} transactions where the initiator intended to sell \coin{BTC} ICO for \coin{XRP}. 
We discover that all three offers were filled by the same account \xrpaddr{rMyronEjVcAdqUvhzx4MaBDwBPSPCrDHYm}, who received the aforementioned \coin{BTC} IOU tokens directly from the issuer's account.
Additional evidence on social media reveals that the IOU issuer's account is held by someone named \texttt{Myrone Bagalay}.\footnote{See \url{https://youtu.be/gVoyCEPvO30} and \url{https://www.twipu.com/MyroneBagalay/tweet/1161288087386894341}.} It becomes obvious that the offer taker's address, starting with \texttt{rMyronE}, must belong to the same person.

By tracing the transaction history of the concerned accounts, we notice that the two offer creators' accounts received their initial \coin{BTC} IOU tokens through payments from the offer taker. 
Furthermore, one offer creator's account, \xrpaddr{rU6m5F9c1eWGKBdLMy1evRwk34HuVc18Wg}, was activated by the offer taker's account. 
Now we can safely assume that all the accounts involved are controlled by that \texttt{Myrone Bagalay}, who issued \coin{BTC} IOU tokens and traded them at arbitrarily determined rates with himself.

What \texttt{Myrone Bagalay} did is completely legitimate within the confines of XRPL. 
In fact, one of the key features of the ledger is the flexibility to establish a closed economy with a limited number of mutually-trusting users who can exchange self-defined assets that are not necessarily acknowledged outside the system. 
However, this makes it challenging to gauge the true value transfer on XRPL, since an IOU token's price---which we proxy by its exchange rate against \coin{XRP}---can be easily inflated or deflated.

Additionally, privately-issued IOU tokens that are never exchanged on the ledger, while seemingly worthless, might be valuable to their transactors after all, should they reach an agreement on those tokens' value off the ledger. However, there is no easy way to assess such value, and we leave the analysis of IOUs to future work.

\point{Summary}
In summary, the throughput on XRPL during our observation period appeared to be fraught with zero-value transactions. 
We learned that both transaction volume and token value on XRPL are highly manipulable. 
One must thus fully understand the underlying measurement approach to correctly interpret the resultant statistics.

%% file: sections/6_discussion.tex
\section{Discussion}
\label{sec:discussion}
In this section, we discuss the results from the previous sections and also answer the research questions presented in Section~\ref{sec:introduction} in light of our results.


\subsection{Interpretation of the Throughput Values}
Overall, we observe that the throughput on EOSIO has been volatile since last November, the throughput on Tezos has been very stable over time, and the throughput on XRPL has been stable in general except during the spam episode.
A common factor between all blockchains is that the current throughput is vastly lower than the alleged capacity even during their utilization peaks, and is on average several orders of magnitude lower.
A similarity between EOSIO and XRPL is that the maximum throughput was reached due to DoS attacks on the network. Indeed, the maximum numbers of transactions on EOSIO is due to the \coin{EIDOS} coin airdrop, while the peak on XRPL was due to the network being spammed with payments.
However, while the spam on XRPL appeared to be anecdotal and lasted for roughly two months, the spam attack on EOSIO is persistent and has continued for over six months to date.
That the increased throughput has different implications for each network.
While on XRPL the consequences of such a spam attack are limited, on EOSIO they forced the network to enter congestion mode, hindering normal usage of the network as transactions become too costly due to the elevated threshold for staking.

Unlike XRPL and EOSIO, Tezos has not seen any spam attack and the level of utilization has been consistent, and relatively low, across time. A majority of the throughput is used for consensus, with most of the spikes in the number of transactions due to baker payments, which are also related to consensus.

\subsection{Revisiting Research Questions}
We now return to the research questions posed in Section~\ref{sec:introduction} and seek to understand better how the different blockchains are used in practice, by attempting to answer them based on the data analysis we perform above.

\point{RQ1: used throughput capacity}
Although the maximum throughput of all blockchains appears vastly lower than the alleged capacity, the situation is not as simple for EOSIO and XRPL. As previously discussed, EOSIO started to be congested because of an airdrop, preventing regular users to use the blockchain normally.
During the attack against XRPL, there were several reports of the network being congested~\cite{AtoZMarkets2019, Tulo2019}, showing that although the claimed capacity was much higher, the {\em actual} capacity might have maxed. Nevertheless, it is yet unclear whether the congestion is mainly due to the suboptimal design of blockchain protocols or the physical constraint of participating nodes' infrastructure. 
On the other hand, Tezos has not yet come close to maximizing its actual capacity.
\point{RQ2: classifying actions}
We made a generalized categorization of transaction types.
Some transaction types are common to all blockchains, such as peer-to-peer transactions and account related transactions, while other types of transactions are inherent to the particularities of the underlying blockchain. While XRPL and Tezos contain easily identifiable action types, making them easy to classify, EOSIO does not have pre-defined action types and classifying actions requires knowledge of the account receiving the action.

\point{RQ3: identifying active blockchain participants}
EOSIO has named accounts which makes it easy to identify participants.
XRPL has optional names, which are registered by the most active players such as exchanges.
Tezos endorsements are often created by bakers, who usually publicize their address and are identifiable. However, there is no easy way to identify participants in peer-to-peer transactions and doing so would require using de-anonymization techniques~\cite{10.1145/2660267.2660379,8802640}.
\point{RQ4: detecting DoS and spam}
The blockchains analyzed are currently under-utilized and when spam occurs, their utilization level increases significantly, as seen in~\autoref{fig:throughput-time}. This makes DoS and spam attacks very easy to detect by simply looking at the transactions, as we saw for EOSIO and XRPL.

\subsection{Transaction Fee Dilemma}
Overall, we have seen that there is a dilemma between having lower transactions fees, which induces spam, or having higher transaction fees, which deters legitimate usage of the network.

One the one side, we have seen that both EOSIO and XRPL have chosen to go with extremely low transaction fees, which in both cases resulted in a very large amount of spam.
On the other side of the spectrum, Ethereum, which has transaction fees based on supply and demand~\cite{Wood2014} has seen a 10-times increase in the fees, mainly because of an increase in the utilization of decentralized finance protocols~\cite{gudgeon2020defi}, making it extremely difficult to use for regular users~\cite{eth-defi-gas}.

There has been work on both sides to improve the current situation but, at the time of writing, no significant progress has been made.
Despite fee structure changes having been proposed in XRPL~\cite{xrp-fees}, concerns are that a fee increase discourages the engagement of legitimate users.
In EOSIO, despite the integration of a new rental market for CPU and RAM~\cite{eos-rental-market}, the current fee structure remains problematic, as the network has now been congested for more than half a year, making it hard to use for regular users.
On the Ethereum side of things, changes in the current pricing system to try to reduce the transaction fees have been proposed~\cite{eip-1559} but are still under discussion.

Overall, for a functional and sustainable blockchain system, it is crucial to find a balanced transaction fee mechanism that can make regular usage of the network affordable while DoS attacks expensive~\cite{DBLP:conf/ndss/0002L20}.

%% file: sections/7_related.tex
\section{Related Work}
\label{sec:related}

Existing literature on transactional patterns and graphs on blockchains has been largely focused on Bitcoin.

Ron et al.~\cite{10.1007/978-3-642-39884-1_2} are among the first to analyze transaction graphs Bitcoin. Using on-chain transaction data with more than~3 million different addresses, the authors find that Mt.~Gox was at the time by far the most used exchange, covering over~80\% of the exchange-related traffic. 

Kondor et al.~\cite{10.1371/journal.pone.0086197} focus on the wealth distribution in Bitcoin and provided an overview of the evolution of various metrics. They find that the Gini coefficient of the balance distribution has increased quite rapidly and show that the wealth distribution in Bitcoin is converging to a power law.

McGinn et al.~\cite{mcginn2016visualizing} focus their work on visualizing Bitcoin transaction patterns. At this point, in 2016, Bitcoin already had more than 300 million addresses, indicating exponential growth over time. The authors propose a visualization which scales well enough to enable pattern searching. Roughly speaking, they present transactions, inputs and outputs as vertices while treating addresses as edges. The authors report that they were able to discover high frequency transactions patterns such as automated laundering operations or denial-of-service attacks.

Ranshous et al.~\cite{10.1007/978-3-319-70278-0_16} extend previous work by using a directed hypergraph to model Bitcoin transactions. They model the transaction as a bipartite hypergraph where edges are in and out amounts of transactions and the two types of vertices are transactions and addresses. Based on this hypergraph, they identify transaction patterns, such as ``short thick band'', a pattern where Bitcoins are received from an exchange, held for a while and sent back to an exchange. Finally, they used different features extracted from the hypergraph, such as the amount of Bitcoin received but also how many times the address appeared in a certain pattern, to train a classifier capable of predicting if a particular address belongs to an exchange.

Di Francesco Maesa et al.~\cite{10.1007/978-3-319-50901-3_59} analyze Bitcoin user graphs to detect unusual behavior. The authors find that discrepancies such as outliers in the in-degree distribution of nodes are often caused by artificial users' behavior. They then introduce the notion of pseudo-spam transactions, which consist of transactions with a single input and multiple outputs where only one has a value higher than a Satoshi, the smallest amount that can be sent in a transaction. They find that approximately~0.5\% of the total number of multi-input multi-output transactions followed such a pattern and that these were often chained.

Several other works also exist about the subject and very often try to leverage some machine learning techniques either to cluster or classify Bitcoin addresses. Monamo et al.~\cite{7802939} attempted to detect anomalies on Bitcoin and show that their approach is able to partly cluster some fraudulent activity on the network. Toyoda et al.~\cite{8254420} focus on classifying Ponzi schemes and related high yield investment programs by applying supervised learning using features related to transaction patterns, such as the number of transactions an address is involved in, or its ratio of pay-in to pay-out. 

More recently, a study of EOSIO decentralized applications (DApps) has been published~\cite{huang2020characterizing}.
The authors analyze the EOSIO blockchain from another angle: they look at the DApps activities and attempt to detect bots and fraudulent activities.
The authors identified thousands of bot accounts as well as real-world attacks, 80 of which have been confirmed by DApp teams.

To the best of our knowledge, this is the first academic work to empirically analyze the transactions of Tezos and XRPL, and the first to compare transactional throughput on these platforms.

%% file: sections/8_conclusion.tex
\section{Conclusions}
\label{sec:conclusion}

We investigate transaction patterns and value transfers on the three major high-throughput blockchains: EOSIO, Tezos, and XRPL. 
Using direct connections with the respective blockchains, we fetch transaction data between~\startdate and \finishdate.
With EOSIO and XRPL, the majority of the transactions exhibit characteristics resembling DoS attacks: on EOSIO,~\empirical{95\%} of the transactions were triggered by the airdrop of a yet valueless token; on XRPL, over~\empirical{94\%}---consistently in different observation periods---of the transactions carry no economic value. 
For Tezos, since transactions per block are largely outnumbered by mandatory endorsements, most of the throughput,~\empirical{76\%} to be exact, is occupied for maintaining consensus.

Furthermore, through several case studies, we present prominent cases of how transactional throughput was used on different blockchains.
Specifically, we show two cases of spam on EOSIO, on-chain governance related transactions on Tezos, as well as payments and exchange offers with zero-value tokens on XRPL.

The bottom line is: the three blockchains studied in this paper demonstrate the capacity to support high levels of throughput; however, the massive potential of those blockchains has thus far not been fully realized for their intended purposes.



%% file: sections/99_acks.tex
\section*{Acknowledgment}
The authors would like to thank the Tezos Foundation for their financial support.

%% file: sections/a_framework.tex
We implement an extensible and reusable measurement framework to facilitate future transactions analysis related research.
Our framework currently supports the three blockchains analyzed in this work, Tezos, EOSIO and XRPL but can easily be extended to support other blockchains.
The core of the software is implemented in Go and is designed to work well on a single machine with many cores.
The framework frontend is provided as a cross-platform static binary command line tool.

While the framework is responsible for the heavy lifting and processing the gigabytes of data, we also provide a companion tool implemented in Python to generate plots and tables from the data generated by the framework.
\point{Data fetching}
The framework currently allows to fetch data either using RPC over HTTP or websockets.
Tezos and EOSIO both use the HTTP adapter to retrieve data while XRPL uses the websocket interface.
The data is retrieved from publicly available archive nodes but the framework can be configured to use other nodes if necessary.
The retrieved data is stored in a gzipped JSON Lines format where each line corresponds to a block. Blocks are stored in chunk of $n$ blocks per file --- where $n$ can be configured --- making parallel processing straightforward. It took less than two days to fetch all the data presented in~\autoref{tab:data-summary}.

\begin{figure}[h!]
\begin{lstlisting}[language=json]
{
  "Pattern": "/data/eos_blocks-*.jsonl.gz",
  "StartBlock": 82152667,
  "EndBlock": 118286375,
  "Processors": [{
      "Name": "TransactionsCount",
      "Type": "count-transactions"
    }, {
      "Name": "GroupedActionsOverTime",
      "Type": "group-actions-over-time",
      "Params": {
        "By": "receiver",
        "Duration": "6h"
      }
    }, {
      "Name": "ActionsByName",
      "Type": "group-actions",
      "Params": {
        "By": "name"
      }
    }
}
\end{lstlisting}
  \caption{Configuration file for our measurement framework}
  \label{lis:framework-config}
\end{figure}

\point{Data processing}
The framework provides several processors which can mainly be used to aggregate the data either over time, or over certain properties such as the sender of a transaction.
The framework is configured using a single JSON file, containing the configuration for the data to be processed as well as the specification of what type of statistics should be collected from the dataset. We show a sample configuration file in~\autoref{lis:framework-config}.
This configuration computes three statistics from block~82,152,667 to block~118,286,375, using the data contained in all the files matching~\lstinline{/data/eos_blocks-*.jsonl.gz}.
The framework will compute the total number of transactions, the number of actions grouped using their receiver over a period of 6 hours, and finally the total number of actions grouped by their name.
All the statistics described above can be used for all the blockchains but the framework also supports blockchain-specific statistics where needed.
New statistics can easily be added to the framework by implementing a common interface.

Our framework was able to analyze the data and output all the statistics required for this paper in less than 4 hours using a powerful 48 core machine.

\begin{figure}[ht]
\begin{lstlisting}[language=go]
type Blockchain interface {
    FetchData(filepath string,
              start, end uint64) error
    ParseBlock(rawLine []byte) (Block, error)
    EmptyBlock() Block
}

type Block interface {
    Number() uint64
    TransactionsCount() int
    Time() time.Time
    ListActions() []Action
}

type Action interface {
    Sender() string
    Receiver() string
    Name() string
}
\end{lstlisting}
  \caption{Main interfaces of our measurement framework}
  \label{lis:framework-interfaces}
\end{figure}

\point{Extending to other blockchains}
The framework has been made as generic as possible to allow integrating other blockchains to perform similar kind of analysis.
In particular, the framework contains three main interfaces shown in~\autoref{lis:framework-interfaces}.
The \lstinline{FetchData} method can be implemented by reusing the HTTP or websocket adapters provided by the framework while the \lstinline{Block} and \lstinline{Action} interfaces typically involves defining the schema of the block or action of the blockchain implemented.
In our implementation, adding a blockchain takes on average 105 new lines of Go code not including tests.